\documentclass[sigconf,10pt]{acmart}

\usepackage{amsmath}
\usepackage{algorithm}
\usepackage{algorithmicx}
\usepackage{algpseudocode}
\usepackage{graphicx}
\usepackage{lipsum}
\usepackage{enumerate}
\usepackage{enumitem}
\usepackage{multirow, makecell}
\usepackage{nccmath}
\usepackage{ragged2e}
\usepackage{pifont}
\usepackage[caption=false,font=normalsize]{subfig}
\usepackage{threeparttable} 
\usepackage{xspace}
\usepackage{bbding}
\usepackage{tabularx}
\usepackage{hyperref}
\usepackage{hyperxmp}

\AtBeginDocument{%
  \providecommand\BibTeX{{%
    \normalfont B\kern-0.5em{\scshape i\kern-0.25em b}\kern-0.8em\TeX}}}

\copyrightyear{2026}
\acmYear{2026}
\setcopyright{cc}
\setcctype{by}
\acmConference[MobiCom '26]{The 32nd Annual International Conference on Mobile Computing and Networking}{October 26--30, 2026}{Austin, TX, USA}
\acmBooktitle{The 32nd Annual International Conference on Mobile Computing and Networking (MobiCom '26), October 26--30, 2026, Austin, TX, USA}
\acmPrice{}
\acmDOI{10.1145/3795866.3796690}
\acmISBN{979-8-4007-2505-0/2026/10}

\newcommand{\ie}{\emph{i.e.},\xspace}

\newcommand{\eg}{\emph{e.g.},\xspace}
\newcommand{\etc}{\emph{etc.}\xspace}

\newcommand\figref[1]{Fig.~\ref{#1}}

\newcommand\tabref[1]{Tab.~\ref{#1}}

\newcommand\secref[1]{Sec.~\ref{#1}}

\newcommand{\fakeparagraph}[1]{\vspace{1mm}\noindent\textbf{#1.}}

\newcommand{\sysname}{{\sf AppFlow}\xspace}
\newcommand{\sysnameposs}{{\sf AppFlow's}\xspace}

\ifodd 1

\newcommand\lxc[1]{\textcolor{black}{#1}}

\else

\newcommand\lxc[1]{#1}

\fi

\begin{document}

\title{AppFlow: Memory Scheduling for Cold Launch of Large Apps on Mobile and Vehicle Systems}
%

\author{Xiaochen Li}
\affiliation{Northwestern Polytechnical University\country{}}
\email{xiaochenli@mail.nwpu.edu.cn}

\author{Sicong Liu}
\authornote{Corresponding author: scliu@nwpu.edu.cn}
\affiliation{Northwestern Polytechnical University\country{}}
\email{scliu@nwpu.edu.cn}

\author{Bin Guo}
\affiliation{Northwestern Polytechnical University\country{}}
\email{guob@nwpu.edu.cn}

\author{Yu Ouyang}
\affiliation{Northwest University\country{China}}
\email{ouyangyu@stumail.nwu.edu.cn}

\author{Fengmin Wu}
\affiliation{Northwestern Polytechnical University\country{}}
\email{fenny@mail.nwpu.edu.cn}

\author{Yuan Xu}
\affiliation{Northwestern Polytechnical University\country{}}
\email{npuyuan@mail.nwpu.edu.cn}

\author{Zhiwen Yu}
\affiliation{Harbin Engineering University, Northwestern Polytechnical University\country{}}
\email{zhiwenyu@nwpu.edu.cn}

\begin{abstract}
GB-scale large apps like on-device LLMs and rich media editors are becoming the next-generation trend, but their heavy memory and I/O demands, especially during multitasking, cause devices to reclaim or kill processes, turning warm apps into cold launches. 
The challenge lies not in storing them, but in fast, accurate launching.
For users, 1s is the usability cliff, yet our measurements show 86.6\% of GB-scale cold launches exceed it.
Also, Android Vitals flags only $\geq$5s as slow, exposing a large satisfaction gap.
Existing optimizations are designed in isolation and conflict.
For example, preloading reduces I/O stalls but consumes scarce memory and is undone by reclamation, while reclamation and killing free memory but sacrifice background survivability, leading to repeated cold relaunches. 
Our key insight is that, although multitasking makes runtime behavior complex, each app’s file access pattern remains predictable. 
The challenge lies in exploiting this predictability, \ie preloading without exhausting memory, reclaiming without undoing gains, and killing selectively to preserve background survivability.
We introduce \sysname, a prediction-based system-wide scheduler that integrates a Selective File Preloader, an Adaptive Memory Reclaimer, and a Context-Aware Process Killer. 
Implemented across the Android framework and Linux kernel without app changes, \sysname cuts GB-scale cold-launch latency by 66.5\% (\eg 2s→690ms) and sustains 95\% of launches within 1s over a 100-day test, significantly improving responsiveness and multitasking experience.
\end{abstract}




\maketitle

%

\section{Introduction}
\label{sec:intro}

GB-scale large apps have become the next-generation trend on mobile devices, placing high pressure on memory and I/O resources. 
Today’s popular apps
such as LLM-powered assistants~\cite{GoogleAIEdgeGallery2025}, image-to-text generators~\cite{collective-ai-appstore}, rich-media editors like CapCut~\cite{capcut2025}, streaming platforms such as TikTok~\cite{TikTok2025}, and 3D mobile games like PUBG Mobile~\cite{pubgmobile}, all ship with GB-scale footprints.
Notably, eight of the ten most-downloaded apps on Google Play now exceed 1GB~\cite{appmagic_topcharts,androidauthority_ram2025}. 
Despite limited mobile resources, users typically keep multiple tasks open at once. 
\lxc{For example, \textit{Lucy} may utilize TikTok and RedNote for trip planning, simultaneously leaving an on-device LLM active in the background. This repetitive switching leads to frequent cold launches that break her operational flow. Compounding the issue, the LLM’s intensive, instantaneous memory footprints can cause the system to terminate background apps, resulting in a fragmented and frustrating multitasking experience.}
In in-vehicle systems, drivers cold launch apps like Google Maps for navigation or pedestrian-detection apps for driver assistance, where every second of delay degrades experience and can even affect driving safety.
This common multitasking scenario stretches memory and I/O to their limits.

To launch a GB-scale large app under such pressure, the OS must \textit{reclaim memory} by either swapping cached pages~\cite{lim2023swam,liang2025ariadne} or killing background processes~\cite{AndroidLmkd2025}. 
Both \textit{sacrifice} background app responsiveness: previously warm apps degrade into cold launches. 
When \textit{Lucy} switches back to Instagram, the system must reload more than 1GB of assets from storage, often stalling for over 2s, twice the 1s \textit{usability cliff}~\cite{nielsen1994usability} at which users perceive apps as frozen or crashed.

Crucially, this is not a corner case but a daily inevitability. 
Modern smartphones keep only 20 apps resident~\cite{huang2024more} (even fewer on other devices like vehicles), while users install 80+ on average~\cite{buildfire2025}, making $\frac{3}{4}$ launches cold by design. 
For example, with just 15 GB-scale apps always running (\eg TikTok, RedNote, and PUBG) on a mobile device, \eg Google Pixel 8, 86.6\% of relaunches exceed the 1s usability cliff. 
In short, users inevitably face these “frozen-app moments” multiple times a day (\eg up to 7 times per day in our 100-day tests), breaking the seamless launch in multitasking experience.

Despite extensive efforts, reducing cold launch latency without sacrificing background survivability remains fundamentally hard. 
Prior work tackles only one side of the bottleneck: I/O preloading~\cite{ryu2023fast,garg2024crossprefetch} reduces stalls but consumes scarce RAM and is quickly undone by aggressive reclamation, while memory reclamation~\cite{lim2023swam,li2025pmr} frees space but often evicts preloaded pages, nullifying I/O gains. 
The core difficulty is that preloading and reclamation/killing are designed in isolation and inherently conflict. 
To our knowledge, there is no prior work that jointly optimizes them.
To address these, we face two challenges:

\noindent$\bullet$ \textbf{\textit{Challenge \#1: }}
Deciding when, what, and how to \textit{preload} for balancing cold-launch latency with multitasking is non-trivial. 
GB-scale launches involve \textit{thousands of small}, \textit{scattered} reads that fragment I/O and keep I/O throughput far below peak (see \secref{sec:opportunity:preload}). 
Prior efforts either preload \textit{all} files~\cite{parate2013practical,lee2016context,lee2017cas}, which is prohibitive for GB-scale apps as it exhausts memory and causes reclamation or killing, or load \textit{on demand}~\cite{joo2011fast,ryu2023fast,garg2024crossprefetch}, which saves memory but often lags behind CPU execution, turning file I/O into the bottleneck and making GB-scale cold launches even slower.
Moreover, preloading competes with reclamation for I/O bandwidth and its extra footprint sacrifices background survivability.

\noindent$\bullet$ \textbf{\textit{Challenge \#2: }}
It is intractable to \textit{promptly} reclaim memory for GB-scale cold launches \textit{without breaking background survivability} or \textit{undoing preloading gains}. 
In practice, cold launches stall allocations and even block preloading, yet existing solutions such as Android's LMK~\cite{AndroidLmkd2025} or follow-up studies~\cite{li2025pmr,liang2020acclaim,lim2023swam} explore to address this by reclaiming pages~\cite{liang2020acclaim,li2025pmr} or killing processes~\cite{AndroidLmkd2025}.
However, both of these approaches fall short as they treat \textit{file-backed} and \textit{anonymous} pages \textit{uniformly} despite asymmetric costs~\cite{liang2020acclaim}. 
Also, they inherently trade off new app responsiveness for background survivability, leaving the problem unresolved. 
 
To address these, we present \sysname, a system-wide scheduler that \emph{jointly} manages preloading, page reclaim, and process killing on commodity mobile and vehicle devices.

\noindent$\bullet$ \textbf{\textit{First}}.
We profile GB-scale launches and identify a structural I/O imbalance, \ie although small files are only 3\% of total size, they dominate launch latency by interrupting large-file transfers and fragmenting sequential reads (see \secref{sec:opportunity:preload}). 
Leveraging this, we design a \textit{Selective File Preloader} that decides when, what, and how to preload. 
It preloads small files before launch to minimize memory use and avoid early stalls, while streaming large files during launch with large blocks to maximize throughput. 
A ms-level knapsack solver further selects per-app cutoffs (see \secref{sec:tech:preloader}). 

\noindent$\bullet$ \textbf{\textit{Second}}. 
We \textit{decouple} file-backed and anonymous \textit{page reclamation} to exploit their asymmetric costs: file-backed pages can be reclaimed quickly in large contiguous blocks, while anonymous pages require slow write-back (\secref{sec:oppo:reclaim}). 
Harnessing this, \sysname integrates an \textit{Adaptive Memory Reclaimer} that prioritizes file-backed pages under pressure, switches to anonymous pages otherwise, and tags preloaded data to prevent premature eviction (see \secref{sec:tech:reclaim}). 
It further includes a \textit{Context-Aware App Killer} that removes only the long-running background apps (see \secref{sec:tech:killer}).

We implement \sysname spanning the Android framework and Linux kernel layers, requiring no modifications to application layer. 
We conduct extensive evaluations on three platforms, including commercial smartphones and in-vehicle systems, using 60+ popular apps with diverse workloads. 
Compared with Android OS and three state-of-the-art baselines, \sysname cuts GB-scale cold-launch latency by up to 66.5\% (\eg 2s→690ms). 
Unlike prior systems that sacrifice background responsiveness, \sysname sustains 95\% of launches within 1s over a 100-day test, even under 1.29$\times$ higher concurrency. 
These gains come from 2.35$\times$ higher I/O throughput and 67.9\% lower memory pressure (measured by the number of kernel direct reclaims).
Our main contributions are summarized as follows.

\begin{itemize}
    \item To our knowledge, this is the first to jointly optimize GB-scale cold launch and multitask survivability.
    It achieves sub-second latency (the usability cliff) for 95\% launches and relaunches.
    \item We design \sysname, a system-wide scheduling framework that leverages the predictable app access patterns to unify selective file preloading, adaptive memory reclamation, and context-aware process killing. 
    \item Experiments show that \sysname significantly reduces GB-scale cold launch latency and improves background survivability with higher I/O throughput and more efficient reclamation, compared to state-of-the-arts.
\end{itemize}

\section{Background and Motivation}
\subsection{Primer of App Launch Processes}


Fast app launch is essential for user experience~\cite{son2021asap,shin2012understanding,li2022smartphone,shen2019deepapp}, especially for GB‑scale, memory‑hungry apps (\eg mobile games, LLM‑powered apps) that suffer significant cold launch latency from hardware I/O and memory constraints. 
We note that prior studies~\cite{nielsen1994usability} show that \textit{app launch delays over 1s} is a \textbf{usability cliff}, leading users to suspect network or system failures.
To systematically analyze app launch bottlenecks, we first summarize its fundamental mechanisms, costs, and the underlying hardware and OS limitations.

\begin{figure}[t]
    \centering    \includegraphics[width=0.5\textwidth]{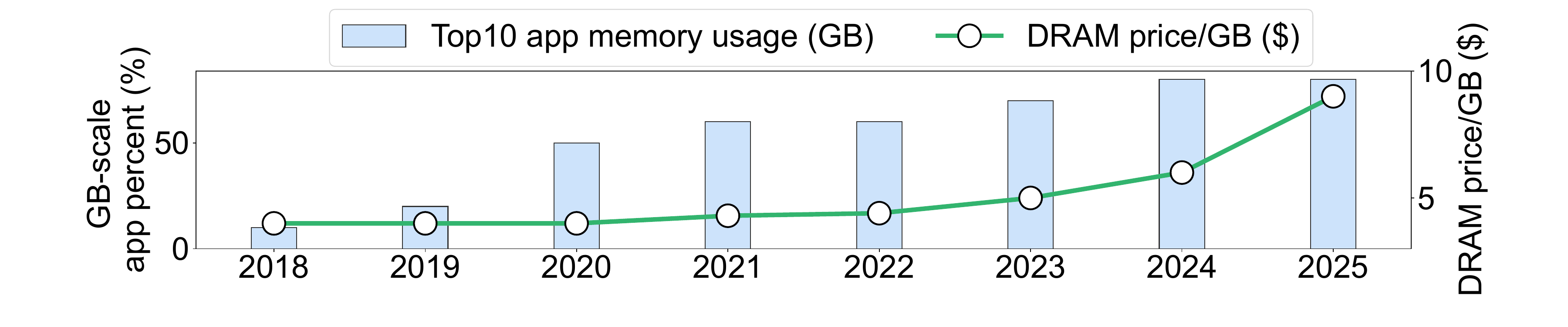} 
    \vspace{-8mm}
    \caption{\lxc{Over the past few years, the share of GB-scale apps in the Google Play top 10 has grown markedly, alongside a rapid surge in DRAM prices.}}
\label{fig:background:memprice}
\vspace{-4mm}
\end{figure}

\subsubsection{The Trend of GB‑Scale Large Mobile Apps}
\label{sec:back:memory-app}
\lxc{Over the past ten years, apps on the Google Play Store have grown markedly in size~\cite{reddit_android_memory_footprint,wikipedia_android_version_history,androidauthority_ram2025}.}
Eight of the top-10 popular apps exceed 1GB~\cite{appmagic_topcharts,androidauthority_ram2025}, and about 30\% of the top 100 apps are over 1GB. 
These GB‑scale apps include \textit{three} main types. 
\textit{i)} \textit{On‑device AI apps}, such as LLM‑powered assistants~\cite{GoogleAIEdgeGallery2025} and image‑to‑text generators~\cite{collective-ai-appstore}, run AI models locally for private and low‑latency inference, driving up RAM usage.
\textit{ii)} \textit{Rich media apps}, \eg video editors CapCut~\cite{capcut2025} and streaming platforms TikTok~\cite{TikTok2025}, embed high‑resolution and multimodal inputs that push memory footprints.  
\textit{iii)} \textit{Mobile games}, \eg graphically intensive titles like PUBG Mobile that load large 3D assets, tend to incur substantial memory footprints. 



\subsubsection{GB-scale Large App Cold-Launch Problems}
\label{sec:back:launch}

\begin{figure}[t]
    \centering    \includegraphics[width=0.48\textwidth]{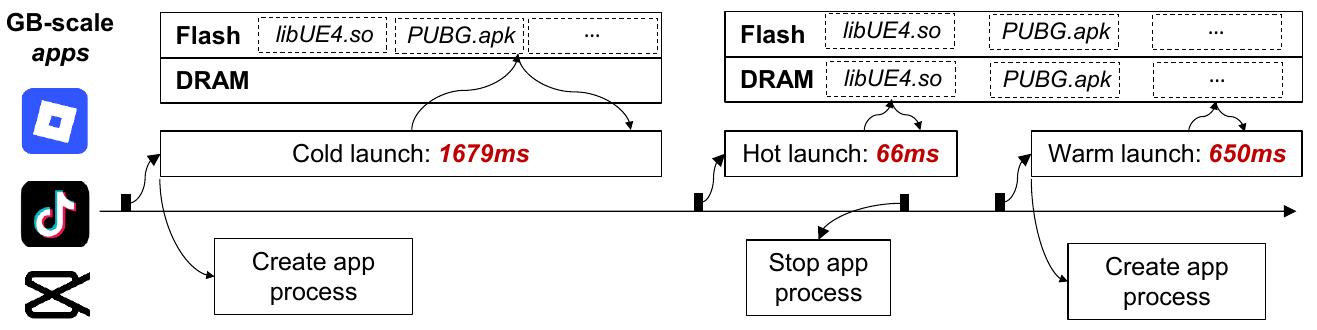} 
    \vspace{-5mm}
    \caption{Illustration of GB-scale app launch progress.}
\label{fig:background:launch_types}
\vspace{-2mm}
\end{figure}

\begin{figure}[t]
    \centering    \includegraphics[width=0.49\textwidth]{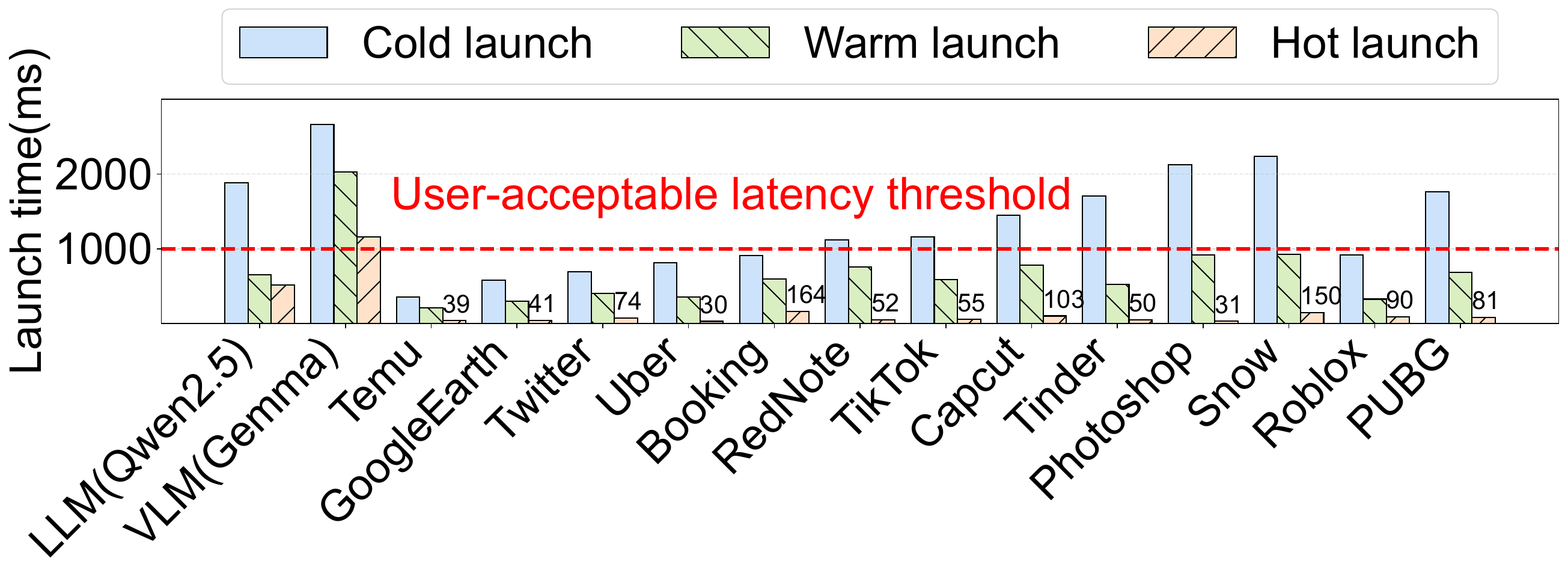} 
    \vspace{-3mm}
    \caption{Illustration of launch latency across cold, warm, and hot modes for 15 popular GB-scale apps.}
\label{fig:background:cold_warm_hot}
\vspace{-4mm}
\end{figure}

Mobile apps typically launch in three modes, \ie \textit{cold}, \textit{warm}, and \textit{hot} launch, and each mode feels different to the user. 
As shown in \figref{fig:background:launch_types}, in a \textit{hot launch}, the app is already in RAM with all its data ready, so it launches almost instantly. 
A \textit{warm launch} happens when the app process has been stopped but its cache remains; it launches more slowly than a hot launch but still fairly quickly~\cite{garg2024crossprefetch}. 
In a \textit{cold launch}, nothing is in memory, so the system must load the APK file (
\eg \texttt{PUBG.apk}) and shared libraries (\eg \texttt{libUE4}, \etc) from Flash into DRAM, making it the slowest~\cite{lim2023swam}.
We run tests on a Pixel 8 smartphone using 15 GB-scale apps across three modes and find that 33\% cold launches often take more than 1s (see \figref{fig:background:cold_warm_hot}).

Two main factors drive this delay. 
\textit{First}, cold launch issues dozens of \textit{I/O requests} that contend for limited mobile DRAM bandwidth, slowing each read. 
\textit{Second}, mobile users often run multiple GB-scale apps (\eg video players TikTok~\cite{TikTok2025}, or LLM‑based assistants Google AI Edge Gallery~\cite{GoogleAIEdgeGallery2025}) until RAM is full or the battery runs out. 
To ensure memory for a new app, Android/Linux either \textit{kills background processes}~\cite{AndroidLmkd2025} or \textit{reclaims cached pages}~\cite{gorman2004understanding}, turning potential hot launches into warm or cold ones.
\lxc{The persistent rise in DRAM prices~\cite{Jeronimo2025GlobalMemoryShortageCrisis,carbone2022_dram_price,trendforce_dram_spot_2025} poses a significant constraint on memory expansion (see \figref{fig:background:memprice}), making it difficult to increase the number of apps that can be kept in the active state for hot launches.}
Existing study~\cite{huang2024more} shows mobile devices can keep about 20 apps active, yet the typical user has around 80 apps installed~\cite{buildfire2025}, resulting in a cold launch for 75\% apps, highlighting how \textit{often} cold launches occur and their impact.

Moreover, frequent multitask switch makes cold launch delays more problematic. 
For example, if a user switch from TikTok to Line and back, both apps load instantly only if they remain in RAM, otherwise each load is a slow cold launch.
Several systems~\cite{huang2024more,liang2025ariadne,lebeck2020end} use aggressive memory management strategies to \textit{maximize the number of hot-launch} apps.
However, typical mobile DRAM holds just three GB‑scale apps at once. 
Keeping state and caches for too many large apps and background services exceeds this limit.

\subsubsection{Mobile Memory Management.}
\label{sec:back:MMM}
Efficient memory management is critical to \textit{securing memory space} (\ie DRAM) for new launches and reducing cold‑start delays of GB‑scale apps on constrained mobile devices.
Mobile OS typically employ two memory management strategies: \textit{i) memory reclamation}~\cite{liang2020acclaim} (\textit{discarding} file‑backed pages and \textit{swapping} anonymous pages to free RAM) and \textit{ii) process killing}~\cite{lim2023swam} (terminating background apps).

\begin{figure}[t]
    \centering    \includegraphics[width=0.46\textwidth]{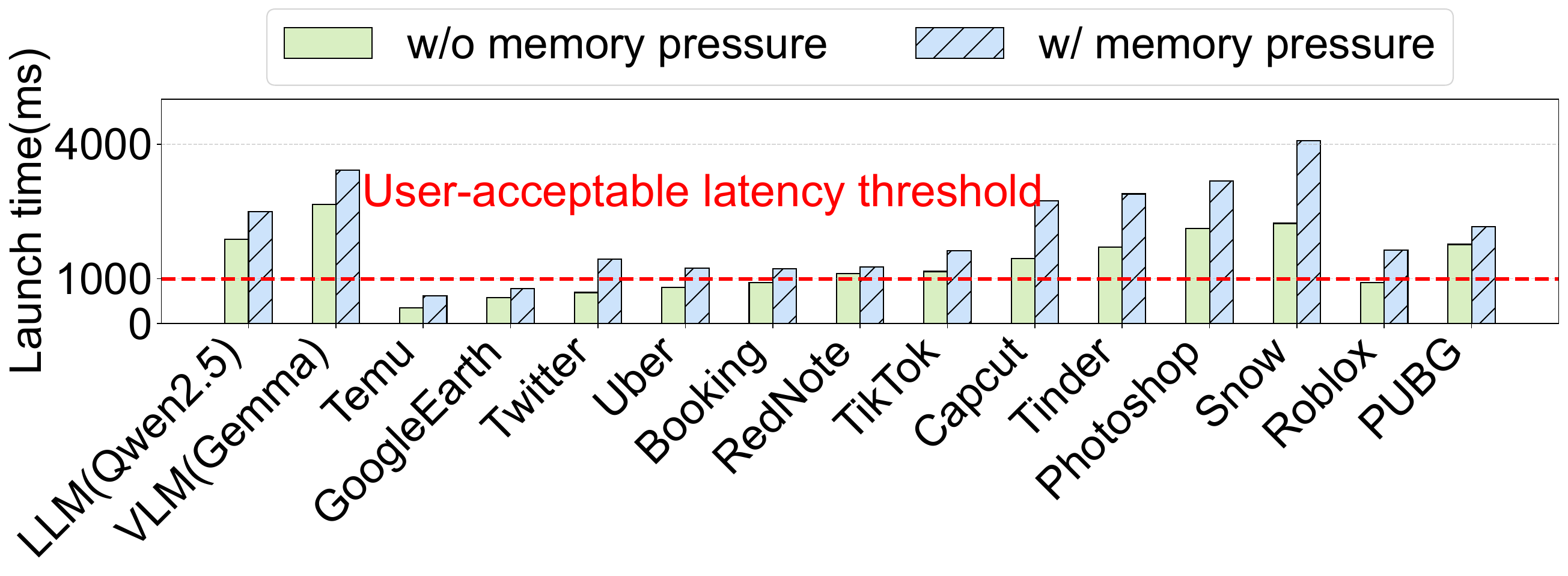} 
    \vspace{-4mm}
    \caption{Impact of memory availability on app launch time: comparing cold‑launch latencies with 10 background apps (<1GB free memory) \textit{vs.} no background apps (>2GB free).}
\label{fig:background:cold_cold}
\vspace{-2mm}
\end{figure}

Specifically, when free RAM drops below a threshold, the kernel’s \textit{reclamation} process (\texttt{kswapd}) runs a two‑phase cleanup:  
\textit{i) discard file‑backed pages} (code/resources) that can be reloaded later~\cite{son2021asap}.  
\textit{ii) swap anonymous pages} to the Flash before freeing RAM~\cite{kim2017application,liang2022cachesifter,li2024elasticzram}.  
However, writing to flash is slow. 
Limited I/O bandwidth on mobile devices always causes reclamation to lag behind allocation, especially under high memory pressure.  
In our test (\figref{fig:background:cold_cold}), launching a new app with ten apps in the background takes up to 2.2$\times$ longer than with no background apps, causing 86.6\% of launches to exceed 1s.
%
If reclamation and swap can’t free enough RAM, due to limited swap space or heavy memory pressure, the OS will \textit{kill} background processes to reclaim their entire working sets. 
While this ensures memory space for the new app, it forces killed running apps into \textit{cold launches} on next use, degrading user experience.
%
To further reduce cold‑launch delays, \textit{file preloading}~\cite{ryu2023fast,parate2013practical,yan2012fast} is an orthogonal technique.
By loading critical binaries and data into memory before~\cite{parate2013practical} or during launch~\cite{ryu2023fast}, the system can overlap I/O with other tasks and shorten launch time.
However, it is non-trivial to speed up memory reclaim for fast cold launches while keeping background apps hot.

\subsection{Observation}
\label{sec:motivation:opp}
This subsection presents our key observation, revealing opportunities for optimization.

\begin{figure}[t]
    \centering    \includegraphics[width=0.46\textwidth]{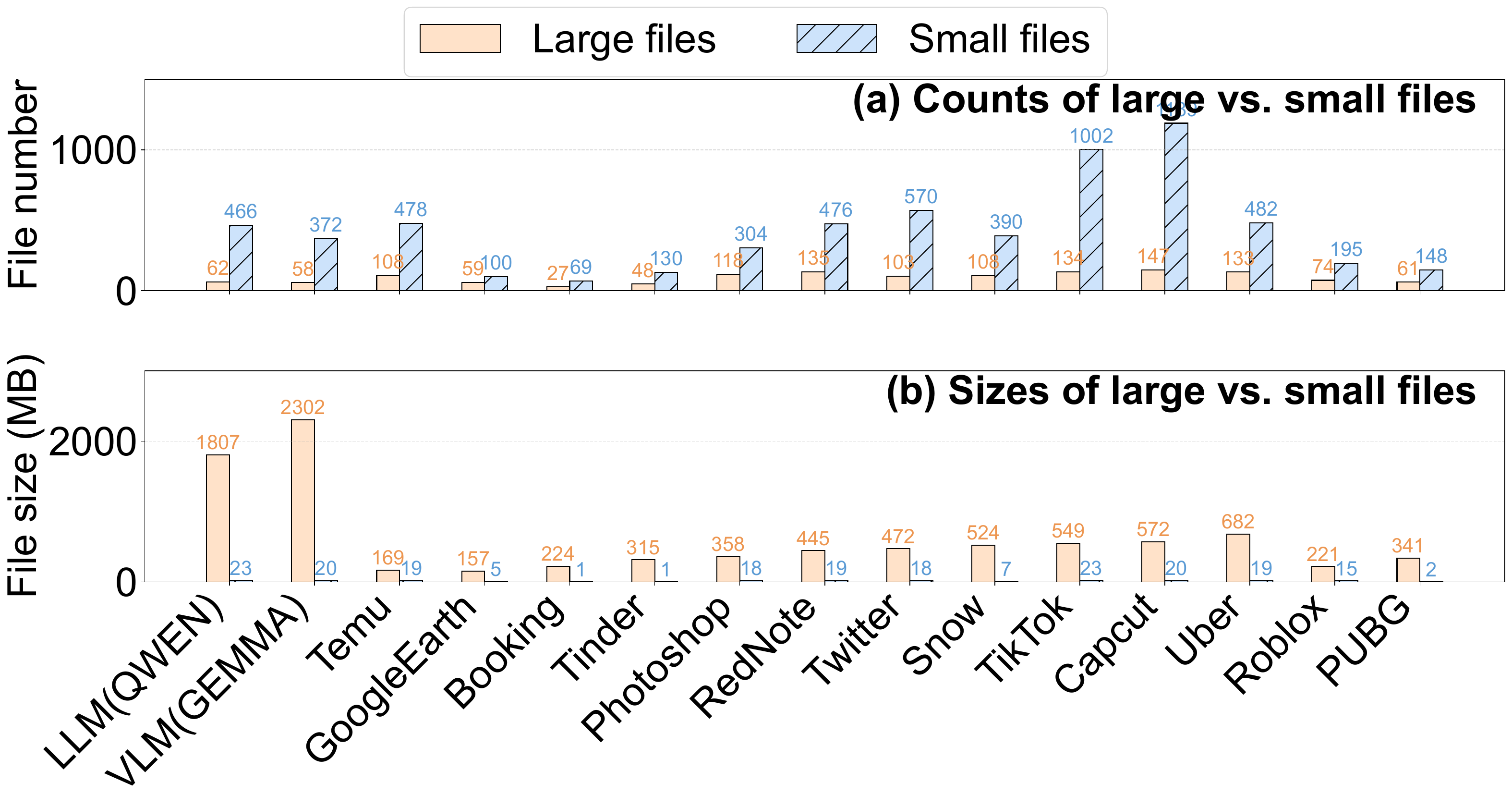} 
    \vspace{-4mm}
    \caption{During GB-scale large app launch, small-file($\leq$128 KB) accesses outnumber large-file($>$128 KB) accesses by 4.7$\times$ yet occupy just 2.2\% of their memory.}
\label{fig:oppo:preload:cnt_size}
\vspace{-2mm}
\end{figure}

\begin{figure}[t]
    \centering    \includegraphics[width=0.46\textwidth]{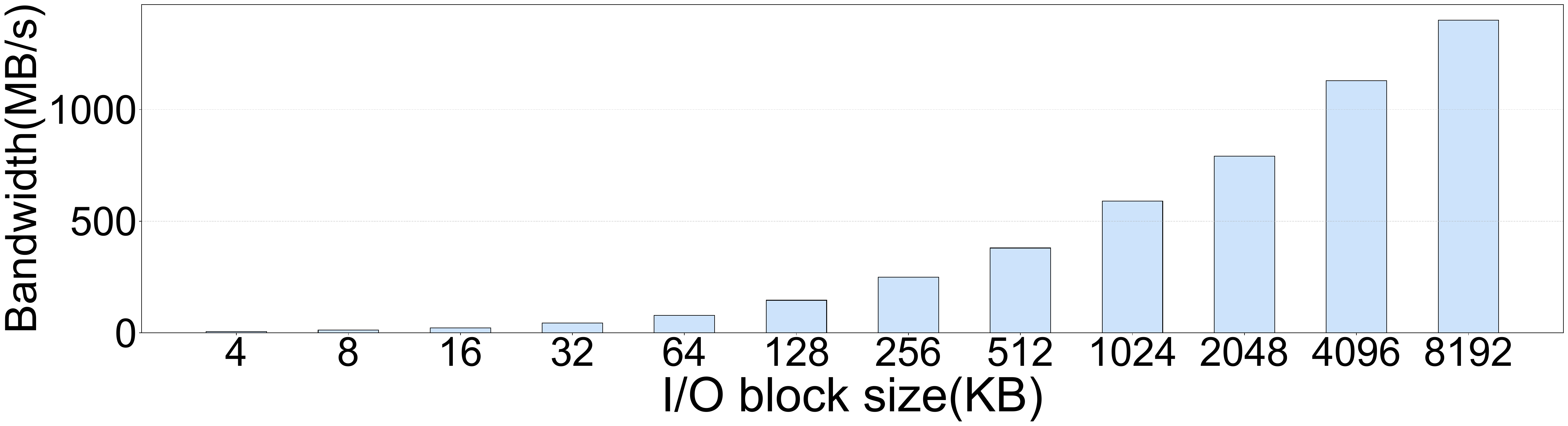} 
    \vspace{-3mm}
    \caption{I/O bandwidth varies with  I/O block sizes.}
\label{fig:obs:IO_speed}
\vspace{-2mm}
\end{figure}

  

\subsubsection{Idle I/O and Hot-set File}
\label{sec:opportunity:preload}
As mentioned above, efficient preloading can hide cold launch I/O latency.
Towards it, we have three findings.
\lxc{\textit{First}, we empirically find that 97.2\% of cold‑launch I/O waits across 24 apps involve \textit{predictable} static data (\eg UE4~\cite{sanders2016introduction} engine assets or LLM weights). 
Crucially, we observe that 79\% of DRAM I/O bandwidth remains \textit{unused} during app switches. 
This indicates the existence of \textit{idle windows} that are ideal for preloading.}
\textit{Second}, we also profile fifteen GB-scale
apps over ten launches each, the results show that GB‑scale apps actually \textit{depend on many small hot-set files} whose total size is small. 
\lxc{For example, as shown in \figref{fig:oppo:preload:cnt_size}, 1,002 small files are accessed during the TikTok~\cite{TikTok2025} launch, which is 7.47$\times$ the number of large files.} 
While they occupy only 23MB of memory, which is just 4\% of the memory used by large files.
\textit{Third}, scaling up I/O blocks boost throughput for big files but waste bandwidth on small ones, while scaling down blocks suit tiny files yet limit throughput and add latency. 
Tuning I/O block size in a structured manner can boost I/O throughput significantly (see \figref{fig:obs:IO_speed}). 


\begin{figure}[t]
    \centering    \includegraphics[width=0.48\textwidth]{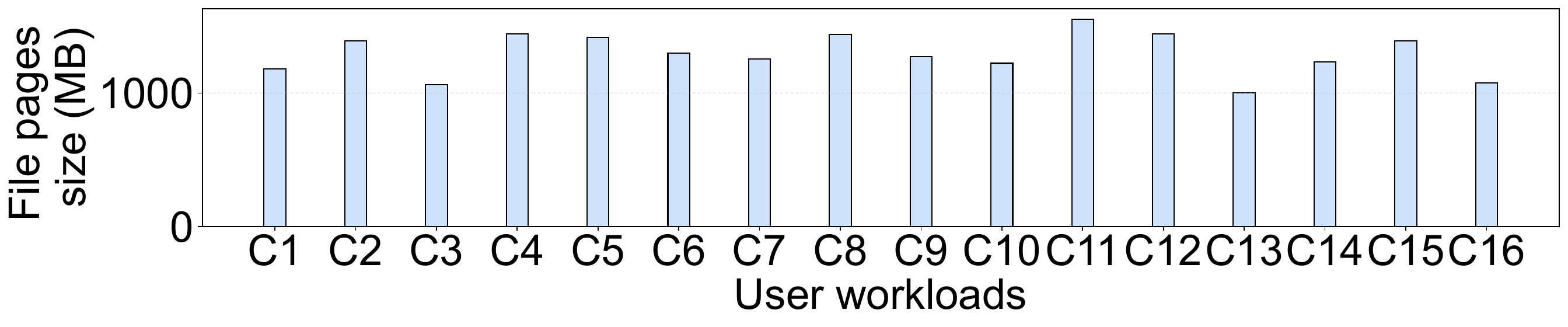}
    \vspace{-8mm}
    \caption{Diverse file page sizes across 16 representative user workloads~\cite{tian2020identifying}. 
    For example, C1 is a meal booking scenario, typically occurring in the afternoon, where users launch both social apps like WhatsApp and media apps like TikTok.}
\label{fig:obs:filepages}
\vspace{-3mm}
\end{figure}

\subsubsection{Page-varying Reclamation Cost}
\label{sec:oppo:reclaim}
Reclamation cost varies by page type.
\textit{Anonymous pages} are expensive to reclaim, as they require I/O writes to Flash (\eg \texttt{swap\_writepage()}), while \textit{file-backed pages} are cheap to reclaim, as they can be dropped without writes.
Existing reclamation strategies alternate \textit{evenly} between them, but this interleaving forces the system to wait for expensive writes before freeing cheap pages.
\textit{Moreover}, our analysis of 16 typical user workloads~\cite{tian2020identifying} (see \figref{fig:obs:filepages}) reclaimable file‑backed pages often exceed 1000 MB, regardless of nonlinear growth in social apps or linear growth in media apps.

\textbf{Observation}: Different page types require distinct reclamation strategies. 
Dropping file-backed pages \textit{first} under high memory pressure frees memory instantly, avoiding I/O delays. 
Anonymous page writebacks can \textit{then} be handled after peak demand, ensuring faster memory reclamation.


\subsubsection{Temporal Killing Impact}
\label{sec:obser_kill}
When memory reclamation alone cannot free enough RAM, the OS’s low‑memory killer (LMK) may terminate background processes to secure memory space. 
However, this blunt approach causes problems.
\textit{First}, killing apps that users may re-open quickly disrupts workflow.
Existing user behavior studies report that 92.5\% of apps are re-accessed within 30 minutes~\cite{tian2020identifying}, yet LMK still kills 62\% of these high‑probability apps, causing duplicate I/O and expensive cold relaunches.
\textit{Second}, mobile users tend to focus on certain app categories at diverse time (\eg productivity tools in the morning, media apps in the evening). 

\textbf{Observation}: Integrating the \textit{temporal locality} into kill decisions can significantly cut unnecessary kills, saving I/O and reducing user discomfort.

\subsubsection{Preload–Reclaim Interference}
\label{sec:oppo:preload}

\begin{figure}[t]
  \centering
  \subfloat[Preload I/O throughput.]{
    \includegraphics[height=0.11\textwidth]{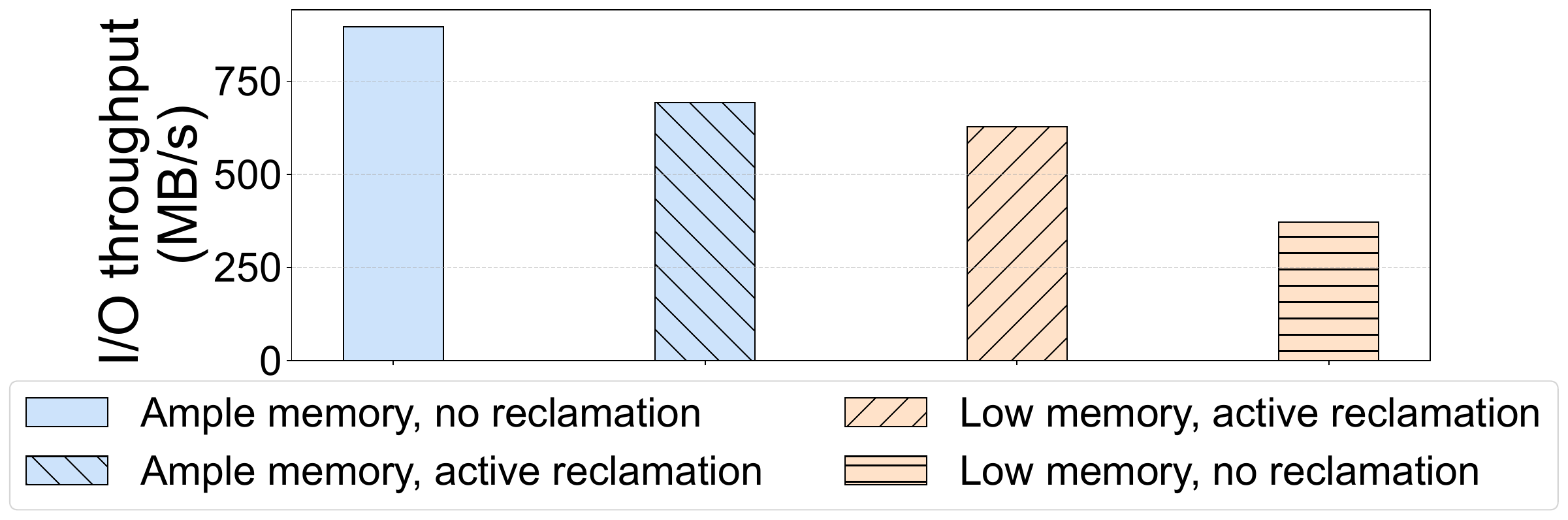}
    \label{fig:oppo:io:throughput}
  }
  \subfloat[I/O latency.]{
    \includegraphics[height=0.11\textwidth]{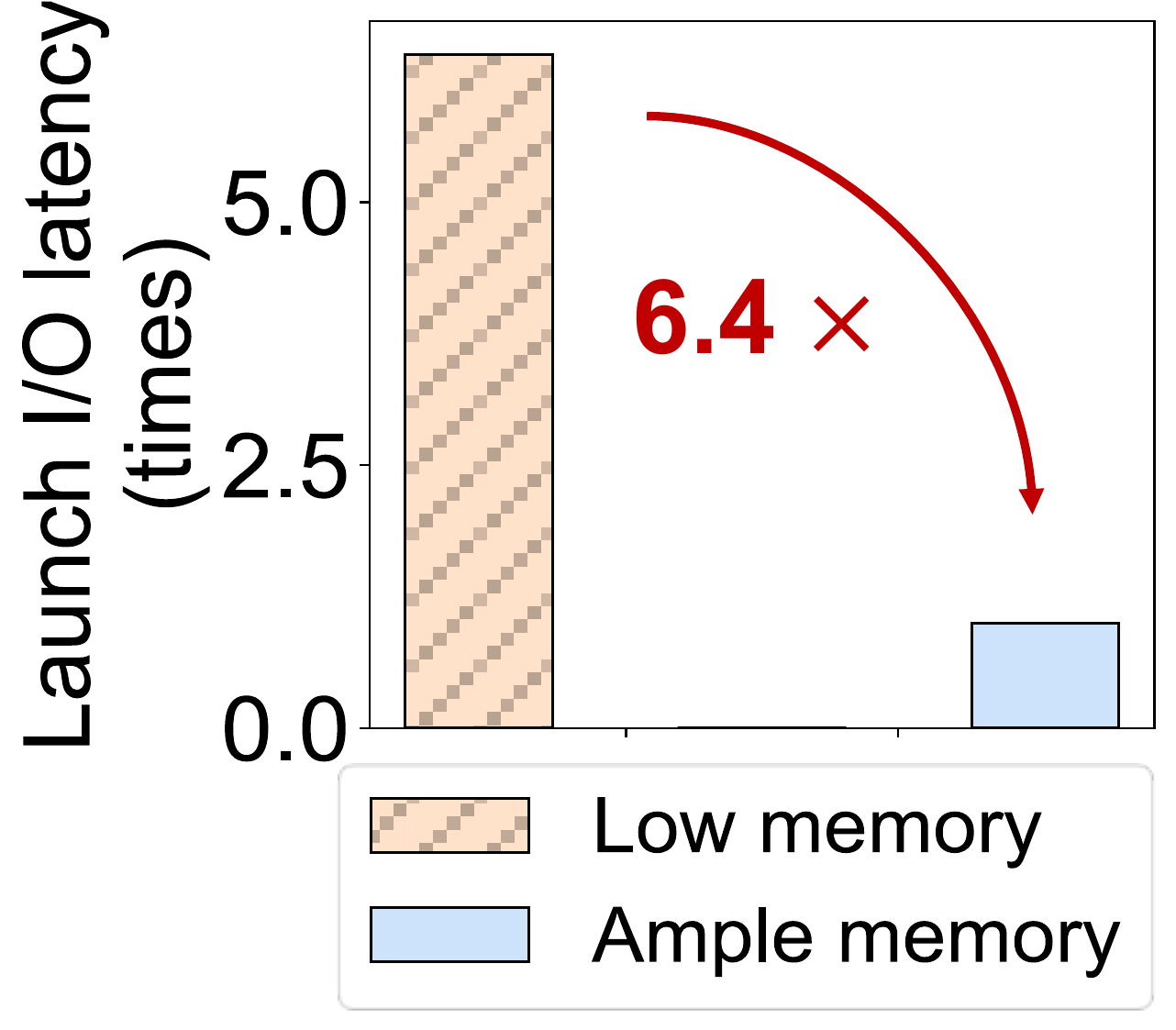}
    \label{fig:oppo:io:pgfault}
  }
    \vspace{-2mm}
  \caption{Analysis of preload-reclaim interference. (a) memory reclamation reduces the throughput of preloading;
(b) reclaiming preloaded files results in a 6.4$\times$ longer I/O latency during GB-scale app launch.}
  \label{fig:oppo:Preload-reclaim}
  \vspace{-4mm}
\end{figure}
Cold launch delays for GB‑scale apps arise from two tightly coupled \textit{bottlenecks}, \ie \textit{limited I/O bandwidth} slows critical file reads, and \textit{low free memory} blocks both new allocations and preloading. 
Speeding up cold launch means loading files and freeing RAM at the same time, yet these two actions often compete and cut each other’s gains. 
It mainly comes from two areas:
\begin{itemize}
    \item \textbf{I/O contention}: When reclaiming anonymous pages, the reclamation increases I/O write traffic that competes with preload reads.
    \item \textbf{Page eviction}: When reclaiming file-backed pages, the reclamation may evict preloaded pages, forcing apps to re-read them during launch.
\end{itemize}

To quantify their interference, we \textit{first} test I/O throughput under four memory conditions: \textit{i) }ample memory, no reclamation; \textit{ii) }ample memory, active reclamation; \textit{iii) low memory, active reclamation}; and \textit{iv)} low memory, no reclamation (~\figref{fig:oppo:io:throughput}). 
I/O throughput is worst in case (iv) because stalled allocations leave the storage under-utilised. 
Even with ample memory, turning on reclamation (case ii) still cuts throughput as write traffic from anonymous-page reclaim competes with pre-load reads. 
We \textit{next} measure the extra I/O latency after pre-loading to quantify their interference in terms of \textit{page eviction},  under both ample- and low-memory conditions (Fig. \ref{fig:oppo:io:pgfault}).
With low memory, launch-time I/O latency jumps by 6.4$\times$, showing that many preloaded pages are evicted during reclaim and had to be re-read.

\textbf{Observation}: 
Without proper coordination, memory reclaim and pre-loading can negate each other's benefits.
\lxc{This observation suggests that predictable launch-time dependencies are necessary but insufficient. Preloading must be coordinated with reclamation to reduce contention and avoid immediate eviction, motivating a system-wide scheduler that (i) exploits idle I/O windows, (ii) prioritizes hot-set files under memory limits, and (iii) reclaims without negating preloading gains.}

\section{Overview}
\label{sec:overview}

\subsection{Problem Formulation}
\label{sec:problem}

Our goal is to minimize the cold‑launch latency for GB‑scale apps while ensuring \textit{relaunch} of background apps sharing the same I/O and memory resources completes within $1s$ (a known responsiveness threshold~\cite{nielsen1994usability}), \ie every $T_{i} < 1$, to maintain a smooth user experience. 
The cold launch latency $T^{cold}_i$ is decomposed as:

\begin{equation}
\label{equ:time_sum}
    T_{i}^{cold} = T_{i}^{I/O} + T_{i}^{cpu} + T_{i}^{alloc}
\end{equation}
where $T_{i}^{I/O}$ represents the data load time from Flash to DRAM, the primary component of launch time for each app. 
$T_{i}^{cpu}$ is the app initialization time, and $T_{i}^{alloc}$ represents the memory allocation time, which depends on the system's available memory during app launch.
%
Since $T_{i}^{cpu}$ depends on each app’s internal initialization logic, which we cannot modify at the system level, thus we target the reduction of the I/O and memory delays.
On mobile devices, the total memory usage can be expressed as:
\begin{equation}
    M = M^{sys} + M^{aval} + \sum_{i} p_{i} + \sum_{j \in \mathcal{J}} m_{j} 
\end{equation}
where $M^{sys}$ is system usage (\eg the UI rendering engine), and $M^{aval}$ is the available RAM for allocation. 
$p_i$ is the memory preloaded for app $i$ to be launched.
$\mathcal{J}$ is the background apps, and $m_j$ is the memory occupied by each app $j \in \mathcal{J}$.
Assume app $i$ must read $M_i^{file}$ bytes of files and allocate $M_i^{sum}$ bytes of RAM during launch. 
Let $v^{I/O}$ be the I/O throughput and $v^{rec}$ the memory‑reclamation throughput, which depend on I/O block size and available memory, respectively (refer to \secref{sec:motivation:opp}). 
Then the two key sub‑latencies become:
\begin{equation}
    T_{i}^{I/O} = \frac{M_{i}^{file}- p_{i}}{v^{I/O}}
    \label{equ:t_io}
\end{equation}
\begin{equation}
    T_{i}^{alloc} =  \frac{Max\{(M_{i}^{sum}-M^{aval}),0\}}{v^{rec}}
    \label{equ:t_mem}
\end{equation}
Here, $p_i$, $M^{aval}$, $v^{I/O}$, and $v^{rec}$ are the design variables we can tune to improve launch performance.
We define the feasible design space $\mathcal{F}$ by:
\begin{align}
\mathcal{F} = \{(p_i, M^{aval}, v^{I/O}&, v^{rec})\mid \text{C1, C2} \}\\
\text{C1:}\quad M^{sys} + M^{aval} + \sum_{i\in\mathcal{I}}& p_{i} + \sum_{j\in\mathcal{J}} m_{j}
\;\le\; M,\\
\text{C2:}\quad T_{i}^{\text{cpu}} \;+\; T_{i}(p_{i},M^{aval}) \;\le&\; 1s,
\ \forall \ i \in \ app \ set \mathcal{J}.
\end{align}
The combined I/O and allocation latency is:
\begin{equation}
    T_{i}(p_{i},M^{aval}) = \frac{M_{i}^{file} - p_{i}}{v_{I/O}} + \frac{Max\{(M_{i}^{sum}-M^{aval}),0\}}{v_{rec}}
\end{equation}
With the trend of GB‑scale apps on memory‑limited devices (see \secref{sec:back:memory-app}), $C1$ tightly bounds $p_i$ and $M^{aval}$.
Ensuring every running app in $\mathcal{J}$ meets the sub‑1s launch bound under these tight memory limits ($C2$) makes the interactive tuning of $M^{aval}$, $v^{I/O}$, and $v^{rec}$ extremely challenging.

\subsection{Solution Overview}
\begin{figure}[t]
    \centering    \includegraphics[width=0.47\textwidth]{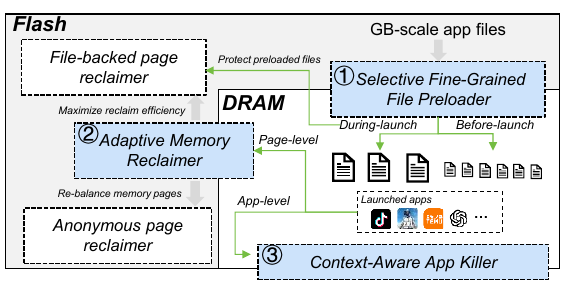} 
    \vspace{-3mm}
    \caption{Workflow of AppFlow.}
\label{fig:tech:overview}
\vspace{-3mm}
\end{figure}

To address above intractable problems, we propose \sysname, a system-wide memory scheduling framework tailored for mobile devices. 
As shown in \figref{fig:tech:overview}, \sysname comprises three tightly-coupled components operating at different granularities and optimization dimensions, \ie preloading, reclamation, and killing, forming a comprehensive solution.
In particular,
the \textit{\textbf{Selective File Preloader}} (file-level) proactively utilizes available free memory and I/O to preload critical file pages, reducing cold-start latency by up to 55.2\% (see \secref{sec:tech:preloader}).
When free memory runs low, the \textit{\textbf{Adaptive Memory Reclaimer}} (page-level) keeps preloader efficiency by reclaiming clean pages from background apps at a rate tuned to real-time I/O load, while tagging and skipping the pages just fetched to avoid wasteful reloads (see \secref{sec:tech:reclaim}).
Also, the proposed Reclaimer is preloader-aware, which deliberately skips pages the Preloader has just loaded, preventing costly reloads.
When proactive trimming still falls short, \sysname activates the \textit{\textbf{Context-Aware App Killer}} (process-level), that goes beyond Android’s LRU list.
It targets long-running, memory-bloated apps whose relaunch footprint returns to baseline, avoiding repeat kills and cutting cold-launch time by 29.8\% in long-term tests (see \secref{sec:exp:case}).
Together, they enable \sysname to maintain fast launch of GB-scale large apps even under tight memory pressure and high concurrency.
As we will show in \secref{sec:experiment}, \sysname reduces cold launch latency by up to 66.5\% across three categories of GB-scale large apps while improving background concurrency by 2.6$\times$. 
Moreover, \sysname sustains 95\% of launches within 1s over a 100-day test, demonstrating that a system-wide memory scheduler that jointly orchestrates preloading, page reclamation, and process killing is critical for mobile responsiveness.

\section{System Design}

\subsection{Selective File Preloader}
\label{sec:tech:preloader}

Modern GB‑scale apps on DRAM‑constrained mobile devices frequently stall on \textit{blocking I/O} during cold launches, such as CPU tasks wait for file reads from flash to DRAM. 
As mentioned in \secref{sec:opportunity:preload}, preloading can fill these gaps, but to avoid excessive memory use, we must carefully decide \textit{when} (timing), \textit{what} (which files), and \textit{how} (scaling I/O block size) to preload. 
By tuning these three dimensions, we can overlap I/O with idle periods.

\begin{figure}[t]
  \centering
    \subfloat[Before-launch preloading]{
    \includegraphics[width=0.48\textwidth]{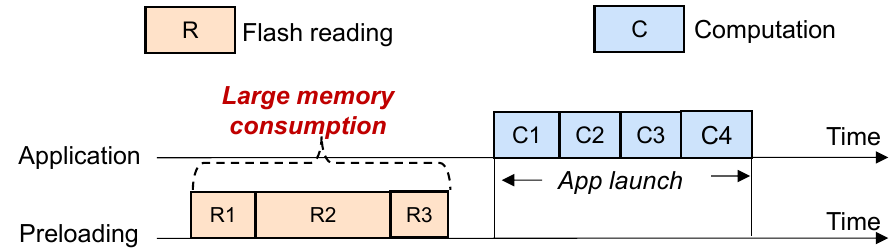}
    \label{fig:tech:preload:predict}
  }\\
  \subfloat[During-launch preloading]{
    \includegraphics[width=0.48\textwidth]{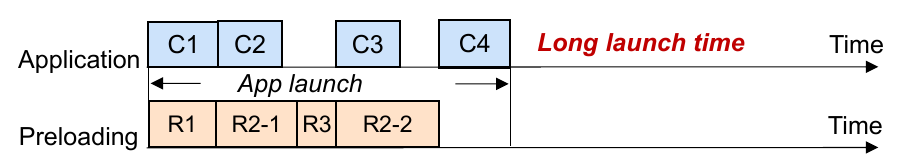}
    \label{fig:tech:preload:demand}
  } 
  \\
  \subfloat[Our selective file preloader]{
    \includegraphics[width=0.48\textwidth]{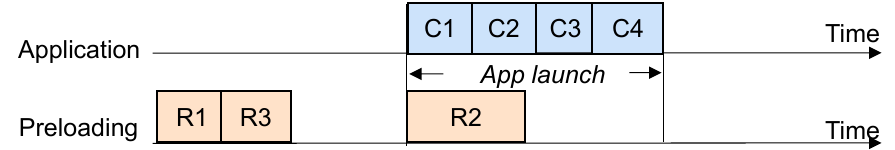}
    \label{fig:tech:preload:proposed}
  }
  \caption{Illustration of diverse preloading methods.}
  \label{fig:background:preloading}
  \vspace{-4mm}
\end{figure}

\textbf{Limitations of Existing Methods}.
Prior efforts struggle to balance launch speed and memory efficiency.
\textit{First}, \textit{preloading before launch} loads all required files in advance~\cite{yan2012fast,hort2021survey,parate2013practical}, eliminating I/O delays but incurring high memory usage, which is often unacceptable on mobile devices. 
For example, as shown in \figref{fig:tech:preload:predict}, preloading just three popular apps, \eg PUBG (0.5 GB), Genshin (0.43 GB), and Gemini (1.8 GB), requires approximately 3GB of memory, while Pixel 8 phone typically allocates only less than 1GB for preloading~\cite{lim2023swam}.
\textit{Second}, \textit{preloading during launch} conserves memory by loading files on demand at runtime, but still suffers from I/O bottlenecks when file access lags behind CPU execution.
As shown in \figref{fig:tech:preload:demand}, the computation steps C3 and C4 are significantly delayed due to the long I/O time of R2 and R3.

These limitations call for a fine-grained approach that aligns preload \textit{timing} with file characteristics.
Actually, GB-scale apps' launch-critical files include two types:
\textit{i) Small but frequent files (\eg <128KB)} contribute minimal memory overhead but cause frequent random I/O if not preloaded early.
\textit{ii) Large and I/O bandwidth-intensive files (\eg $\geq$128KB)}, whose loading time depends on I/O bandwidth. 
Preloading them improves performance, but consumes significant memory.
As discussed in \secref{sec:opportunity:preload}, small, frequent files dominate launch I/O (91\%) in GB\mbox{-}scale apps, and their footprint is modest  ($\approx$3\% of the total; see ~\figref{fig:oppo:preload:cnt_size}).
Moreover, we find that the I/O bandwidth of large files with \textit{large block size} (\ie $\geq$128KB) can be up to 22.8$\times$ faster than with \textit{small block size}. (\figref{fig:obs:IO_speed}).


Based on these, 
to reduce launch latency without exceeding memory limits, we introduce a \textit{time-split}, \textit{file-and-app-type-aware} preloading strategy with two phases. 
The \textit{key idea} is to preload I/O frequency-intensive files \textit{before} launch with small I/O block size, to reduce latency-critical blocking with minimal memory overhead, and preload the remaining large files \textit{during} launch with large I/O block size, to maximize I/O efficiency under time constraints.
Specifically, \sysname employs two complementary strategies:

\noindent$\bullet$
\textit{Before launch, frequency-first}.
This stage focuses on loading as many frequency-intensive small files as possible within a constrained memory budget.
It preloads all small files (<$s_{i}$, decided by each app $i$) and frequently accessed \textit{hot segments} of large files. 
These files add minimal memory overhead but eliminate high-frequency I/O stalls.

\noindent$\bullet$
\textit{During launch, throughput-first}.
Once the app starts, this stage \textit{aggressively} loads large files using \textit{large block size} ($\geq s_{i}$) to maximize I/O throughput of large files. 
However, overlapping with foreground I/O poses a new challenge, \ie I/O contention between the preloader and the app itself.
To address this, we run the preloader as a separate \textit{low-priority}, \textit{separate} process, ensuring that it never blocks or delays foreground I/O requests.

\textit{Selecting} the optimal app-specific cut-off size $s_i$ that separates small ($< s_{i}$) from large  files is necessary yet challenging. 
A large $s_i$ raises prelaunch memory use, while a smaller $s_i$ leaves too many small files unpreloaded, throttling I/O during launch.
We pose $s_i$ selection as a discrete optimization:
\begin{align}
    \text{arg \ min}\, T_{i}^{I/O} &= \frac{M_{i}^{file}-p_{i}}{v_{i}^{I/O}} \nonumber, \quad p_{i} = P(s_{i}), \quad v_{i}^{I/O} = V(s_{i}) 
    \\ s.t.\quad \sum_{i}&p_{i}\leq M_{p} \nonumber
\end{align}
where $P(s_i)$ is the size of files below $s_i$ (preloaded), and $V(s_i)$ is the achieved I/O bandwidth when streaming the remaining files.
Both $P$ and $V$ increase with $s_i$. Since I/O block sizes are powers of two (\ie 4KB, 8KB, \etc), we profile $P$ and $V$ only at these discrete sizes. We cast the selection as a multiple-choice knapsack: each app picks one precomputed $s_i$ under the preload budget $M_p$ (set as 100MB, see in \secref{sec:exp:sens}). The instance is small, solved in $<0.1$ms (see \secref{sec:exp:overhead}); we reoptimize $s_i$ only when an app’s launch profile changes (\eg a GB-scale large app launched), so runtime overhead is negligible and configurations remain near optimal.

\subsection{Adaptive Memory Reclaimer}
\label{sec:tech:reclaim}
Even with preload optimization, \textit{insufficient memory availability} can still block launch, particularly when the system must reclaim memory before fulfilling new allocation requests.
To address this, this module \textit{shifts focus} to reducing the application's \textit{memory allocation time} $T^{alloc}$ while avoiding the degradation of preloading efficiency under mobile memory constraints.

As discussed in \secref{sec:back:launch}, when available memory $M^{aval}$ is sufficient, $T^{alloc}$ is negligible. 
When $M^{aval}$ is low, memory allocation requests during launch may be \textit{blocked} until the system \textit{reclaims} enough memory.
Thus, $T^{alloc}$ is primarily determined by the \textit{memory reclamation speed} $v^{rec}$.
Formally, this can be approximated as
$T^{alloc} = \frac{Max\{(M^{sum}-M^{aval}),0\}}{v^{rec}}$.
Here, $M^{sum}$ is the app’s total memory demand during launch.

\begin{figure}[t]
    \centering \includegraphics[width=0.48\textwidth]{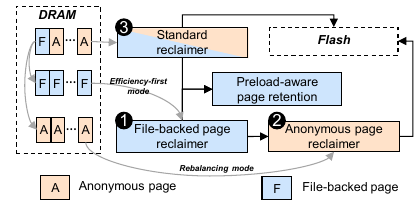}
    \vspace{-2mm}
    \caption{Adaptive memory page reclaimer.}
\label{fig:tech:reclaim}
\vspace{-2mm}
\end{figure}

\textbf{Preliminary of Prior Arts}.
Android/Linux-based mobile systems (\eg smartphones, wearables, robotics) rely on the kernel's background reclaimer thread \texttt{kswapd}, to free memory when $M^{aval}$ is low, during which the app launch is blocked.
\texttt{kswapd} targets \textit{two main types of memory pages}, \ie 
\textit{file-backed pages}, which are mapped from Flash, and \textit{anonymous pages}, which are dynamically generated during runtime.
\texttt{kswapd} \textit{evenly alternates} between these two types, attempting to free roughly balanced file-backed and anonymous pages. 
However, we observe that this is inefficient under high memory pressure due to their fundamental \textit{heterogeneity} and \textit{asymmetry} in reclamation cost (see \secref{sec:oppo:reclaim}).
As shown in \ding{174} of the \figref{fig:tech:reclaim}, evenly alternating reclaim blocks fast file-backed pages reclamation behind slow anonymous page handling.


    

\begin{table}[t]
\tiny
\caption{Anonymous (Anon.) and file-backed (File.) memory sizes (MB) across diverse GB-scale large apps.}
\vspace{-2mm}
\begin{tabular}{|c|c|c|c|c|c|c|c|}
\hline
\textbf{Types}                                                              & \textbf{Apps} & \textbf{Anon.} & \textbf{File.} & \textbf{Types}                                                                  & \textbf{Apps} & \textbf{Anon.} & \textbf{File.} \\ \hline
\multirow{4}{*}{\textbf{\begin{tabular}[c]{@{}c@{}}Multi\\  media\end{tabular}}} & Photoshop     & 636            & 379            & \multirow{2}{*}{\textbf{\begin{tabular}[c]{@{}c@{}}On-device \\ V/LLM\end{tabular}}} & Qwen2.5       & 307            & 1637           \\ \cline{2-4} \cline{6-8} 
                                                                                 & Snow          & 679            & 370            &                                                                                      & Gemma         & 710            & 1925           \\ \cline{2-8} 
                                                                                 & RedNote       & 778            & 403            & \multirow{2}{*}{\textbf{Game}}                                                       & PUBG          & 636            & 392            \\ \cline{2-4} \cline{6-8} 
                                                                                 & Tiktok        & 616            & 434            &                                                                                      & Roblox        & 856            & 281            \\ \hline
\end{tabular}
\label{tab:anon_file}
\vspace{-4mm}
\end{table}

\subsubsection{Page-type-aware Memory Reclamation}
To boost memory reclaim efficiency, we decouple file‑backed and anonymous page reclamation.
During cold launches, we first reclaim only file‑backed pages, freeing RAM instantly without extra I/O or CPU overhead, to boost $v^{\text{rec}}$ and cut allocation latency $T^{\text{alloc}}$. 
However, we observe that since most GB-scale large apps produce on average 1.92$\times$ more anonymous pages than file‑backed pages (see \tabref{tab:anon_file}), favoring file pages alone leads to reclamation starvation, where only anonymous pages remain and allocations stall.

We therefore introduce an \textit{adaptive, page‑type‑aware} strategy that switches modes based on runtime pressure: i) in high‑pressure phases (\eg cold launch), \sysname reclaims only file‑backed pages for maximum speed; ii) otherwise, reclaims anonymous pages to clear costly regions ahead of future bursts. The key challenge is detecting high‑pressure windows without app changes or developer input.

We leverage kernel’s \textit{page‑allocation rate} as an effective, low‑overhead, and app-agnostic signal. 
By instrumenting \texttt{\_alloc\_pages\_nodemask()} with a simple counter polled periodically (\eg every 100ms), we compute $n_{\text{alloc}}$, the pages allocated in the last interval. 

We classify a phase as memory‑sensitive if
$$
M^{aval} < M^F \quad \& \quad n^{alloc} > N^{alloc}
$$
Here, $M^F$ is system‑specific threshold, $N^{\text{alloc}}$ is set as 12800 (\secref{sec:exp:sens}), and $M^{aval}$ is the current free RAM. 
Upon detecting a memory‑intensive condition, \sysname switches to the \textit{efficiency‑first} mode that reclaims only file‑backed pages (\ding{172} in \figref{fig:tech:reclaim}). 
When the pressure subsides, it shifts to a \textit{rebalancing} mode that restores the normal file‑to‑anonymous reclamation ratio (\ding{173} in \figref{fig:tech:reclaim}). 
Finally, \sysname falls back to the standard reclamation method (\ding{174} in \figref{fig:tech:reclaim}).
This fully automated, multi-phase controller runs with negligible overhead and requires no app changes, making it practical for deployment.

\subsubsection{Preload-aware Page Retention}
\label{sec:tech:reclaim:retention}
Reclaiming only file-backed pages during launch can evict preloaded data before it is used.
To keep those pages in RAM until the app actually uses them, we introduce a preload-aware retention strategy that treats the two preload phases differently.
The \textit{key idea} is that each preload phase has different retention needs. 
The files fetched before launch are few and need only a short residency; 
we keep them alive by lightly touching their pages every ten seconds, a scan that takes about 10 ms and adds just 0.1\% overhead. 
Once the GB-scale large app becomes active, those pages are dropped from future scans and can be reclaimed. 
By contrast, the files fetched during launch are far more numerous, making periodic scans too costly. 
Instead, we run a preload-file check in the memory-reclaim path \texttt{shrink\_page\_list}. If a page maps to a preloaded file, the reclaimer skips it and marks it active to avoid rescanning. The Selective File Preloader publishes the preloaded-file list via \texttt{/proc} and clears it after the app launch completes. As a result, preloaded pages remain in DRAM until the launch finishes and are not evicted prematurely.

Algorithm \ref{alg:amr:workflow} shows the pipeline. 
Every 100ms, \sysname reads $n_{\text{alloc}}$ and $M^{aval}$, then chooses the file‑only or full reclamation policy (lines 4).
During the file-backed pages reclamation, \sysname avoids reclaiming the preloaded pages (lines 5).
Once the file-backed page reclamation completes, the anonymous page reclaimer is invoked to rebalance the page type ratio (lines 6).
As we will demonstrate in \secref{sec:exp:contribution}, this adaptive scheme enables both rapid memory recovery and balanced long-term reclamation, preventing anonymous-page buildup and avoiding starvation, thereby reducing launch latency and the frequency of cold launches.

\begin{algorithm}[t]
\caption{Adaptive Memory Reclaimer}
\begin{algorithmic}[1]
\State \textbf{Init}: thresholds $M^{F}$, $N^{alloc}$; \textbf{every 100ms} update $n^{alloc}\gets\textsc{AllocRate}()$.
\For{\textbf{each} allocation event}
  \State $M^{aval}\gets\textsc{FreeMem}()$
  \If{$M^{aval}<M^{F}\ \land\ n^{alloc}>N^{alloc}$}
     \State \textsc{ReclaimFilePages}(\textbf{exclude} protected files)
     \State \textsc{ReclaimAnonPages}()
  \EndIf
\EndFor
\end{algorithmic}
\label{alg:amr:workflow}
\end{algorithm}
\vspace{-3mm}

\begin{figure}[t]
  \centering
  \subfloat[Kill recently-used app]{
\includegraphics[width=0.24\textwidth]{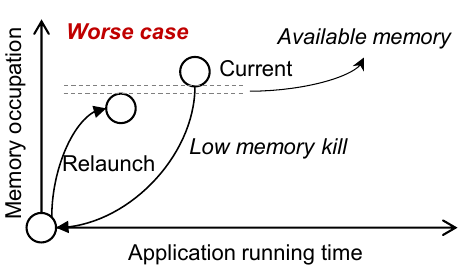}
    \label{fig:tech:kill:worse}
  }
  \subfloat[Kill long-running app]{
\includegraphics[width=0.24\textwidth]{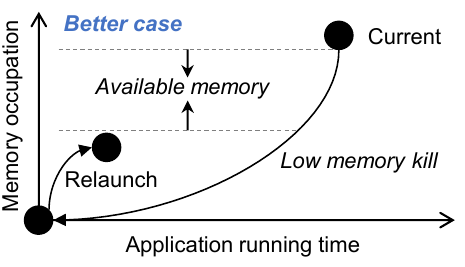}
    \label{fig:tech:kill:better}
  }
\vspace{-3mm}
  \caption{Illustration of diverse killing methods.}
  \label{fig:tech:kill}
  \vspace{-4mm}
\end{figure}

\subsection{Context-Aware Process Killer}
\label{sec:tech:killer}
Even with adaptive page reclamation (\secref{sec:tech:reclaim}), systems sometimes resort to process killing when only anonymous pages remain. 
Two common scenarios cause problematic anonymous‑page dominance, forcing inefficient fallback.

\noindent$\bullet$
\textit{Burst allocations.} 
    During large, repeated allocations, \eg LLM inference, systems defer anonymous page writes (5$\times$ slower than dropping file pages), so anonymous pages accumulate. 
    In our analysis of three categories app traces (\secref{sec:tech:reclaim}), peak anonymous-to-file ratios are 1.92:1, exhausting file pages and leaving only anonymous pages to reclaim.

\noindent$\bullet$
\textit{Reclaim capacity limits.} Anonymous‑page reclamation itself is bounded, either by swap partition size
(\eg 2GB on Pixel 8) or by zRAM capacity (25$\sim$30\% of DRAM). 
Once this bound is reached, anonymous reclamation stalls, and the system falls back to dropping file pages to run out of file pages and trigger process kills.

These scenarios are unavoidable as anonymous pages grow with app activity. 
A clever killer must therefore factor in both the current page‑type imbalance and reclaim capacity limits to avoid unnecessary terminations and preserve both performance and stability.  

\textbf{Limitations of the Kernel LMK}.
When only anonymous pages remain, Android/Linux invokes the low‑memory killer (LMK), terminating apps in background‑to‑foreground order.
As mentioned in \secref{sec:obser_kill}, each kill forces a full cold relaunch, often over 2 seconds for apps like PUBG~\cite{pubgmobile}, then immediately recreates memory pressure (\eg freeing 1200MB but needing 1100MB on relaunch, as shown in \figref{fig:tech:kill:worse}), trapping the system in a kill–restart loop. 
This is because LMK’s blind selection ignores an app’s restart cost, reaccess likelihood, and net memory benefit.

We introduce a novel, \textit{context‑aware app killing} mechanism that leverages two \textit{runtime patterns}, \ie long‑term background bloat and short‑term recency. 
Specifically, we find that \textit{long‑running apps} (\eg social media apps) grow their memory footprint by 30$\sim$50\% over baseline state (\ie the minimal memory used immediately after launch before any runtime cache) yet revert to baseline state on \textit{relaunch} because the process clears all accumulated caches and temporary allocations. 
By harnessing this bloat‑and‑reset phenomenon, \sysname prioritizes reclaiming from these apps to secure large and reliable memory.
Conversely, \textit{recently used apps} pair high reuse probability with elevated memory footprints. 
Rather than kill them following heuristics, we defer termination to avoid the large memory allocation brought by relaunch. 
As shown in \figref{fig:tech:kill:better}, 
\sysname orders kills by net freed memory metric $
\Delta M = M^{\text{current}} - M^{\text{relaunch}},
$ ensuring each kill maximizes lasting memory recovery without penalizing stable apps. 
For recent apps, it switches to \textit{recency‑aware deferment}, overlapping preloading with reclamation idle phases. 
As we will show in \secref{sec:exp:contribution}, this two‑phase strategy reduces 30\% kill events, balancing aggressive memory recovery with seamless interaction.

\section{Implementation}
\label{sec:implementation}
We implement \sysname on Android, one of the most widely used mobile operating systems. 
The Android OS is typically organized into three layers: the application layer, the Android framework, and the Linux kernel. 
\sysname is a memory management mechanism that spans both the Android framework and the Linux kernel. It makes no modifications to the application layer, allowing GB-scale apps to benefit from its acceleration without any code changes.
Specifically, the \textit{Selective File Preloader} and the \textit{Context-Aware Process Killer} modify 1,672 lines of the Android Framework to control preloading and process killing, while the \textit{Adaptive Memory Reclaimer} modifies 1,107 lines of the Linux kernel to manage memory reclamation. 
Since the Adaptive Memory Reclaimer must avoid reclaiming files preloaded by the Selective File Preloader, the two modules communicate via the \texttt{/proc} filesystem.
\section{Experiment}
\label{sec:experiment}

\subsection{Experiment Setup}
\label{subsec_exp_set}

\fakeparagraph{Platforms}
We prototype \sysname on two commercial smartphones (Google Pixel 7 and Pixel 8) and an in-vehicle system powered by the vehicle with a Raspberry Pi 4B handling computation, all running the latest Android 15 operating system (OS)~\cite{android15_about}. 
To emulate memory-constrained low-end mobile devices, we scale the visible DRAM of  both smartphones to 6GB and 8GB by modifying \texttt{arm64\_memblock\_init()}.
And the vehicle DRAM is 4GB by default.

\fakeparagraph{Tasks and workloads}
We evaluate \sysname using both a \textit{real-world 100-day trace} and \textit{simulation workloads}. 
\lxc{The \textit{real-world 100-day dataset}, \sysname-100D, spans weekdays, weekends, and holidays, covering over 60 apps (\eg TikTok, WhatsApp, Camera) and 10,000+ multitask switches, thereby capturing representative user workloads. To collect \sysname-100D, we developed a headless Android application that runs continuously in the background. It utilizes \texttt{UsageStatsManager} to log app-usage sequences for 100 days, after which it automatically terminates and provides a user-facing dashboard allowing participants to review traces before uploading. 
Strict privacy safeguards are enforced: data is locally minimized, retaining only metadata (package names, timestamps, memory states) while strictly excluding all personally identifiable information and app content.
After data collection, we replayed the app switching traces on the same device to evaluate \sysname.}
\textit{Simulation workloads} comprise several GB-scale apps (\eg Gemma, TikTok, Roblox, \etc) and 15 low-memory apps (\eg ChatGPT, Instagram, Zoom, \etc). 
In each test, one GB-scale app is selected for cold-launch under three scenarios: (i) five low-memory apps (\textit{low workload}), (ii) fifteen low-memory apps (\textit{medium workload}), and (iii) fifteen low-memory apps plus two GB-scale apps (\textit{high workload}). \lxc{To ensure consistency in the background workload, we first terminate all background apps and clear the file cache. Next, we employ an automated script to launch the load-generating apps and place them in the background. Once the load is established, we execute the \texttt{am start -W} command to launch the target app and measure its launch time. }

\fakeparagraph{Baselines}
To ensure a comprehensive and fair evaluation, we select four representative baselines that span the design space of preloading and reclamation strategies:

\begin{itemize}
\item \textbf{Android OS}~\cite{android15_about} performs memory reclamation without any preloading. 
This provides the fundamental system baseline.
\item \textbf{Paralfetch}~\cite{ryu2023fast} is a state-of-the-art \textbf{preloading} method that improves launch speed by scheduling I/O sequences to merge small blocks into larger ones. 
\item \textbf{Acclaim}\cite{liang2020acclaim} is a state-of-the-art \textbf{reclamation} method that accelerates memory reclaim by enlarging batch sizes, thereby reducing both the frequency and latency of cold launches under high-load conditions. 
%
\item \textbf{Acclaim + Paralfetch} combines the above SOTA techniques. 
Since, to the best of our knowledge, no joint optimization exists, this baseline demonstrates the limitation of naïve combination.
\end{itemize}

\fakeparagraph{Metrics}
We evaluate \sysname with (i) \textit{user-perceived} metrics, \ie cold-launch latency and frequency of GB-scale apps, captured via \texttt{adb} command~\cite{android_adb}; and (ii) \textit{intermediate} metrics, \ie I/O throughput and memory-pressure events (direct reclaims, LMK), obtained through \texttt{logcat}~\cite{android_logcat} or kernel hooks, to capture both user experience and system behavior.

\subsection{GB-scale App Cold launch performance}
\label{subsec_cold_lunch}
\begin{figure}[t]
  \centering
  \subfloat[on Pixel8-8GB: with low, medium, high workloads]{
    \includegraphics[width=0.49\textwidth]{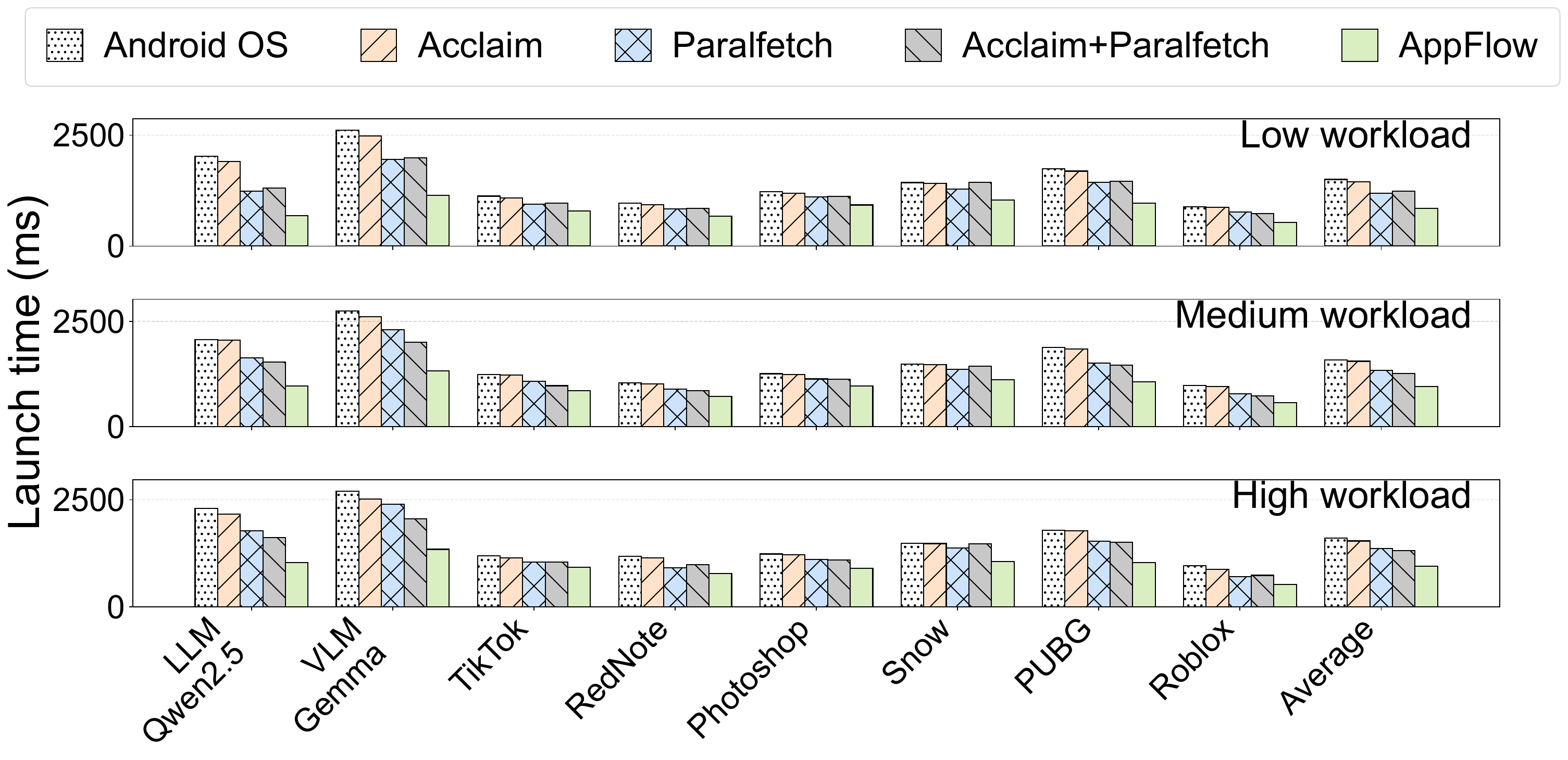}
\label{fig:exp:main:pixel8:8G}
  }
  
  \subfloat[on Pixel8-6GB: with low, medium, high workloads]{
\includegraphics[width=0.48\textwidth]{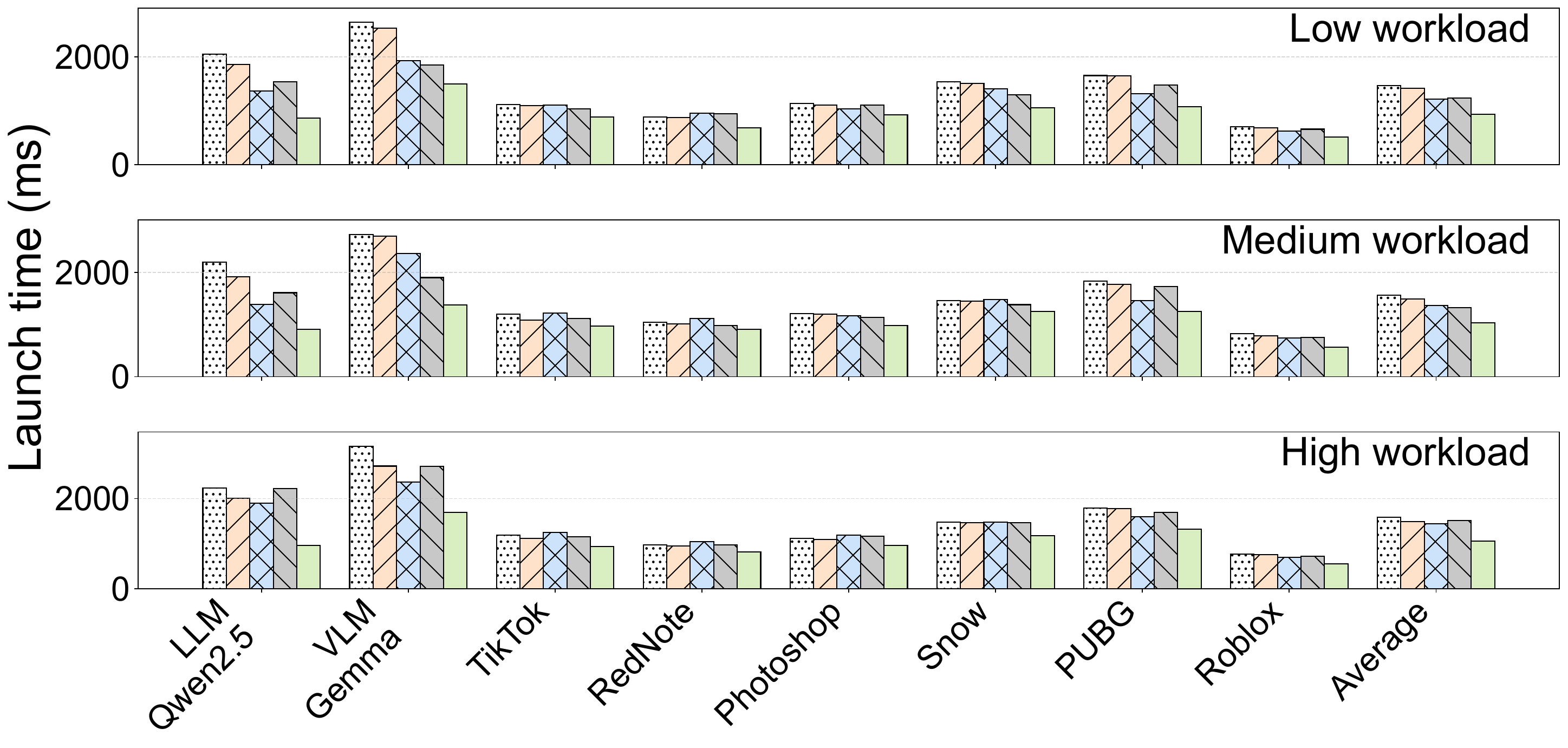}    \label{fig:exp:main:pixel8:6G}
  }
  \caption{GB-scale cold launch latency comparison.}
  \label{fig:exp:main}
  \vspace{-2mm}
\end{figure}
We test the cold-launch time of eight GB-scale apps (\ie LLM-based Qwen2.5-2.0GB, VLM-based Gemma-2.6GB, TikTok-1.1GB, Snow-1.1GB, PUBG-1.2GB, \etc) across four baselines and \sysname on two configurations (\ie Google Pixel 8 devices with 6GB and 8GB limits), each under three workloads (\ie low, medium, high workloads as in \secref{subsec_exp_set}).
\figref{fig:exp:main} shows the results.
\textit{First}, across all device configurations, workloads, and GB-scale apps, \sysname achieves on average 33.7$\sim$43.6\% shorter cold-launch latency than native Android.
\textit{Second}, \sysname outperforms the state-of-the-art preloading baseline Paralfetch even under low workload, reducing cold-launch latency by 29.2\% on average through sustaining higher I/O throughput for GB-scale launches, as shown in \figref{fig:exp:main:pixel8:8G}.
\textit{Third}, \sysname maintains its advantage under memory pressure, reducing launch latency by up to 57\% on a memory-constrained Pixel 8 (6 GB) with high workload (\figref{fig:exp:main:pixel8:6G}).
\textit{Fourth}, naively combining preloading with reclamation may introduce performance degradation; for example, Paralfetch + Acclaim slows Gemma’s launch by 14.4\% over Paralfetch alone and 4.8\% over Acclaim alone under high workload.
\sysname mitigates this by protecting preloaded pages, achieving 56.7\% faster launches than the combination.

\textbf{Summary.} \sysname delivers the \lxc{lowest} cold-launch latency across diverse devices, workloads, and app scales, while avoiding pitfalls of naive baseline combinations, showing robustness and applicability in real-world mobile systems.

\subsection{Multitask Performance}
\begin{figure}[t]
    \centering \includegraphics[width=0.48\textwidth]{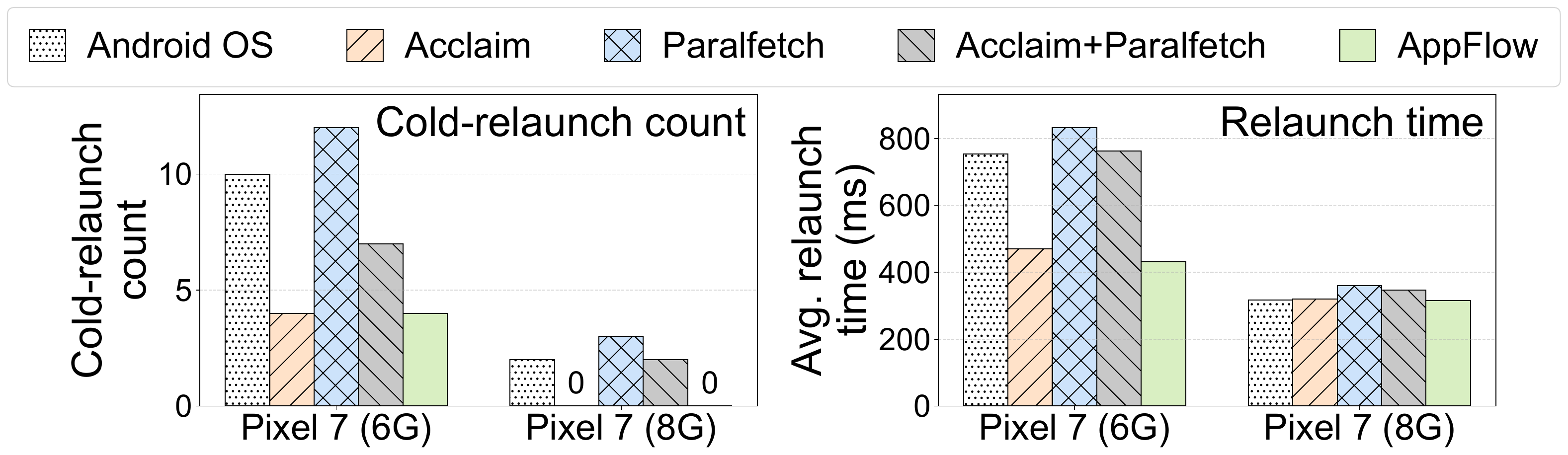} 
    \vspace{-3mm}
    \caption{Multitasking performance with 17 apps: cold-relaunch count (left) and average relaunch time (right).}
\label{fig:exp:concurrent}
\vspace{-4mm}
\end{figure}

We test multitasking performance by comparing \sysname with Android OS~\cite{android15_about}, Paralfetch~\cite{ryu2023fast}, and Paralfetch+Acclaim on Pixel 7 (6/8 GB) under high workload (see \secref{subsec_exp_set}).
Two GB-scale apps(\ie TikTok-1.1GB, Snow-1.1GB) and 15 low-memory apps are each launched twice to simulate multitasking switching from other apps; in the second run, we measure the average relaunch latency and cold relaunch counts. 
\figref{fig:exp:concurrent} shows the results.
\textit{First}, \sysname preserves 1.85$\times$ more background apps (\ie from 7/17 to 13/17) and reduces average launch time by 37.6\% compared to Android OS by rapidly freeing memory at both app and page levels, thus avoiding frequent LMK events. 
\textit{Second}, \sysname prevents the degradation of preloading: on Pixel 7 (6GB), Paralfetch incurs 20\% more cold relaunches than Android OS, while \sysname achieves 60\% fewer. 
This comes from preloading-induced memory pressure that triggers excessive app killing.
\textit{Third}, \sysname incurs 42.8\% fewer cold launches than Paralfetch+Acclaim by protecting preloaded pages while ensuring reclaimed pages are unlikely to reload.

\textbf{Summary.} \sysname sustains multitask responsiveness under high memory pressure, reducing relaunch overheads.

\label{sec:exp:IO}
\begin{figure}[t]
    \centering \includegraphics[width=0.48\textwidth]{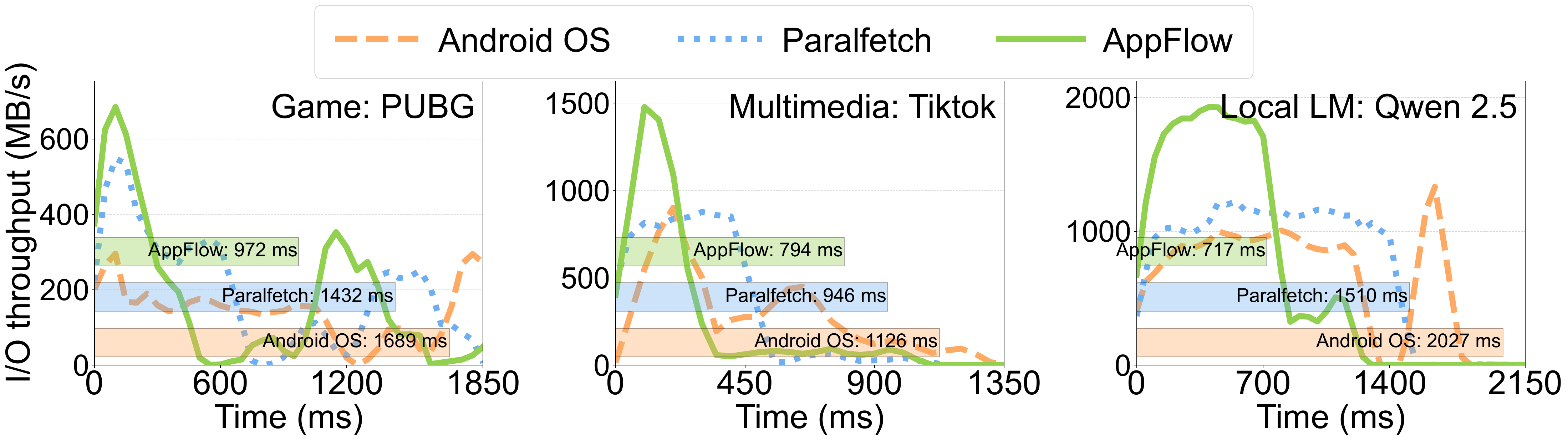} 
    \vspace{-3mm}
    \caption{I/O throughput and cold-launch latency of three GB-scale large apps: PUBG (left), TikTok (middle), Qwen 2.5 (right).}
\label{fig:exp:io}
\vspace{-3mm}
\end{figure}
\subsection{Understanding Performance Gains}
\subsubsection{I/O Throughput During GB-Scale App Launch}
We measure I/O throughput during GB-scale cold launches under a low workload on a Google Pixel 8 (8GB) to illustrate how the \textit{Selective File Preloader} accelerates cold launches (\figref{fig:exp:io}).
\sysname surpasses Android OS~\cite{android15_about} and Paralfetch~\cite{ryu2023fast} via a two-phase, file- and app-aware preloading strategy: small files are prefetched before launch to reduce blocking, while large files are streamed during launch with larger blocks to maximize I/O efficiency. 
In Qwen2.5, it achieves 1.92$\times$ and 1.58$\times$ higher throughput, yielding 29.4\%$\sim$64.6\% faster launches across diverse GB-scale large apps.
\textbf{Summary}. By aligning preloading with file access patterns, \sysname offers a principled path to scalable cold-launch acceleration.

\label{sec:exp:reclaim}
\begin{figure}[t]
    \centering \includegraphics[width=0.48\textwidth]{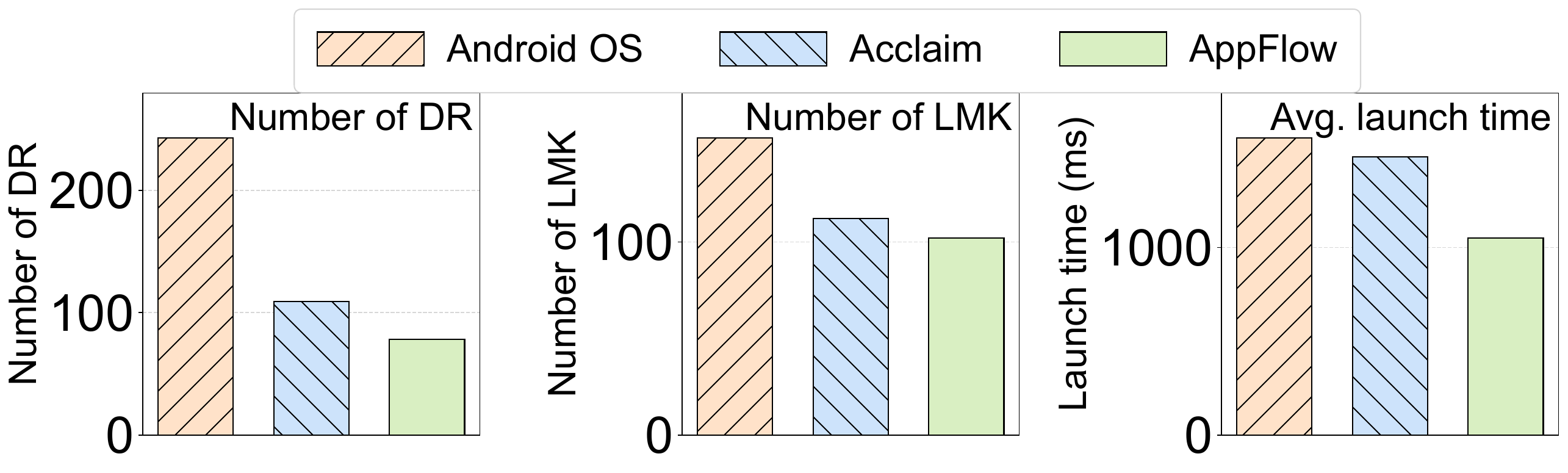} 
    \vspace{-2mm}
    \caption{Memory-pressure mitigation comparison: number of kernel DR (left), LMK events (middle), and average launch latency (right).}
\label{fig:exp:reclaim}
\vspace{-2mm}
\end{figure}

\subsubsection{Memory Reclaim Efficiency During GB-Scale App Launch}
To evaluate how \sysname mitigates memory pressure, we measure kernel direct reclaims (DR), low memory killed (LMK) events, and average launch latency during GB-scale app launches under high workload on Google Pixel 8 (6 GB) (\figref{fig:exp:reclaim}).
\sysname reduces the number of kernel DR by 67.9\% compared to Android OS, cutting allocation stalls and launch delays, and lowers LMK events by 33.7\%, sustaining more background apps. 
These gains come from its Adaptive Memory Reclaimer, which accelerates reclamation by selectively targeting pages, and its Context-Aware App Killer, which prevents repetitive LMKs by prioritizing the termination of memory-intensive apps.
\textbf{Summary}. By combining fine-grained reclamation with intelligent app killing, \sysname turns memory pressure into efficient resource management, preserving responsiveness under GB-scale workloads.

\label{sec:exp:contribution}
\begin{figure}[t]
    \centering \includegraphics[width=0.48\textwidth]{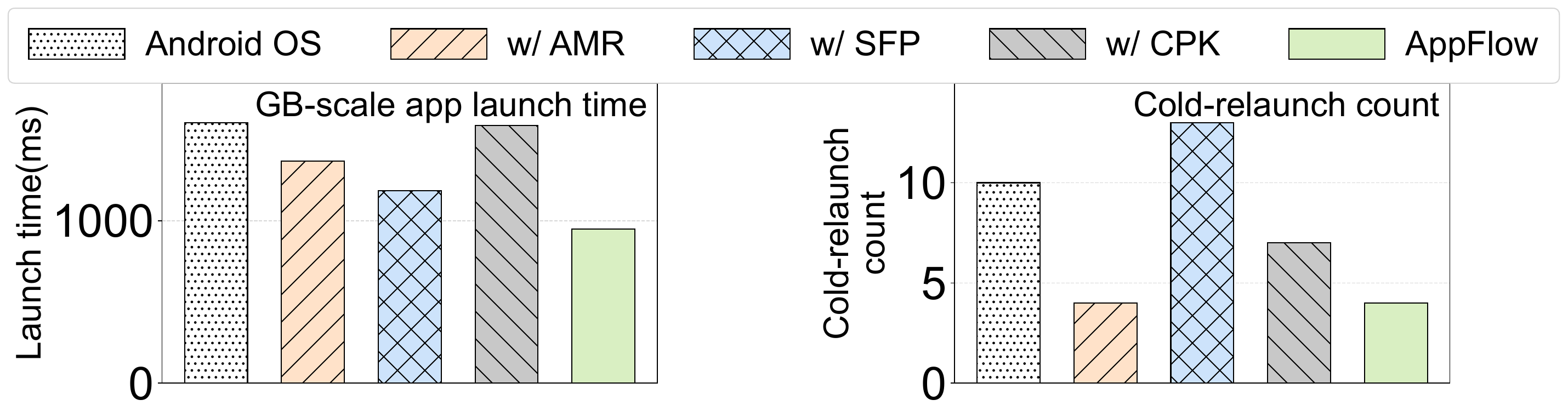} 
    \caption{Ablation study: GB-scale large app launch time (left) and cold-relaunch count(right).} 
\label{fig:exp:ablation}
\vspace{-4mm}
\end{figure}

\subsection{Micro-benchmark}
\subsubsection{Ablation Study}
We evaluate each \sysname module on Pixel 8 (6GB) to quantify its contribution to cold-launch and multitasking performance (\figref{fig:exp:ablation}).
\textit{First}, the Adaptive Memory Reclaimer (AMR) cuts cold launches by 60\% compared to Android OS, though launch time drops only 14.7\%.
\textit{Second}, the Selective File Preloader (SFP) reduces launch time by 26.1\% but increases cold launches by 30\% due to added memory pressure.
\textit{Third}, the Context-Aware Process Killer (CPK) lowers cold launches by 30\% by terminating memory-heavy processes to prevent repeated killings.
\textit{Fourth}. When integrated, these three modules complement each other, delivering a 40.8\% reduction in cold-launch latency while sustaining 1.85$\times$ more background apps.

\label{sec:exp:sens}
\begin{figure}[t]
  \centering
  \subfloat[Preloading size $M^{p}$]{
    \includegraphics[width=0.24\textwidth]{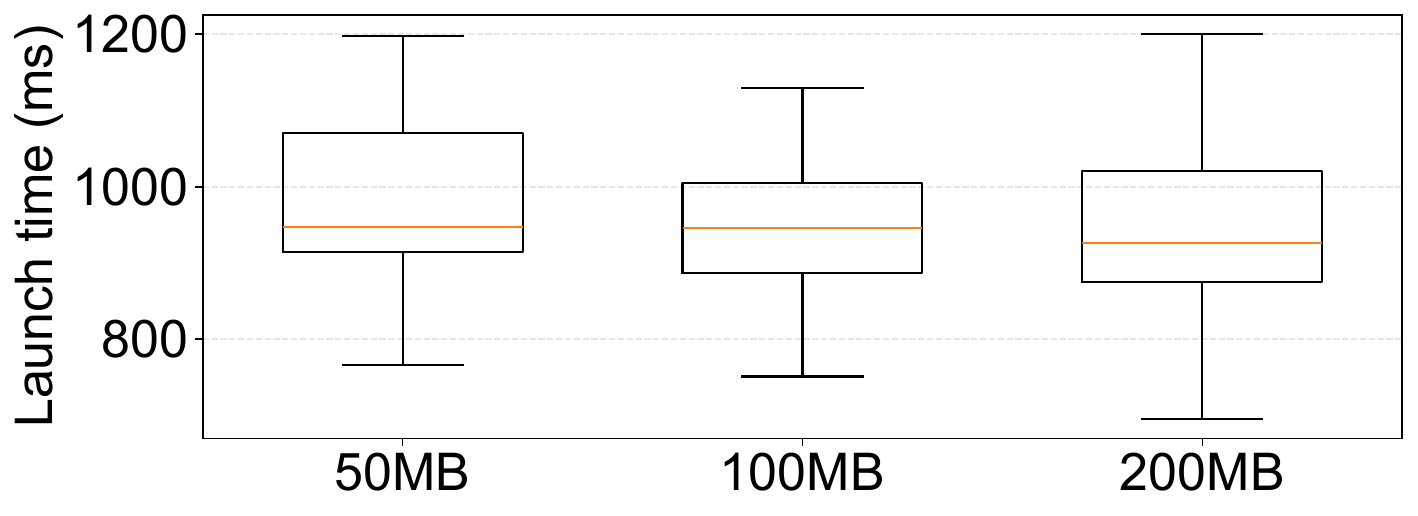}
    \label{fig:exp:sens:preload}
  }
  \subfloat[Allocate thresh. $N^{alloc}$]{
    \includegraphics[width=0.24\textwidth]{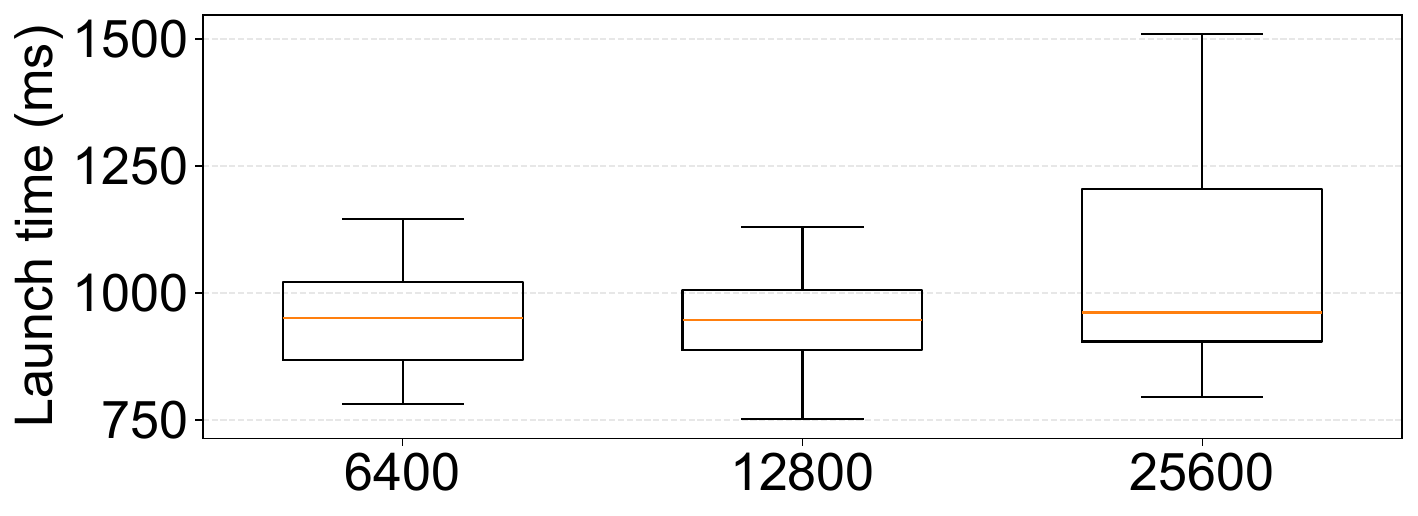}
    \label{fig:exp:sens:amr}
  }
  \caption{Impact of hyperparameter in AppFlow.}
  \label{fig:exp:sens}
  \vspace{-4mm}
\end{figure}

\subsubsection{Hyperparameter Settings}
We explore different hyperparameter settings in \sysname. 
\textit{First, before-launch size $M^{p}$}. 
This controls the amount of data preloaded before launch (\secref{sec:tech:preloader}). 
As shown in \figref{fig:exp:sens:preload}, $M^{p}=100$ MB strikes the best balance. Smaller values fail to cover critical small files (6.5\% slower), while larger values waste memory and increase launch-time variance.
\textit{Second, allocate threshold $N^{alloc}$}. 
It determines when the Adaptive Memory Reclaimer is triggered (\secref{sec:tech:reclaim}). 
As shown in \figref{fig:exp:sens:amr}, $N^{alloc}=12800$ yields the best result; larger thresholds delay reclamation and cause up to 19.9\% slowdown during launch.

\subsubsection{Extra Overhead Analysis}
\label{sec:exp:overhead}
Despite its benefits, \sysname introduces two extra overheads, \ie memory for before-launch preloading and CPU time for computing preload sizes.
\textit{First, memory}. Preloading is capped at 100 MB (<5\% of DRAM on 4 GB devices), and these pages are reclaimed under pressure, leaving performance unaffected.
\textit{Second, CPU time} occurs only when GB-scale large apps change state (minutes apart); even with 60 apps, computation finishes in <0.1 ms, which is also negligible in practice.

\begin{figure}[t]
    \centering \includegraphics[width=0.48\textwidth]{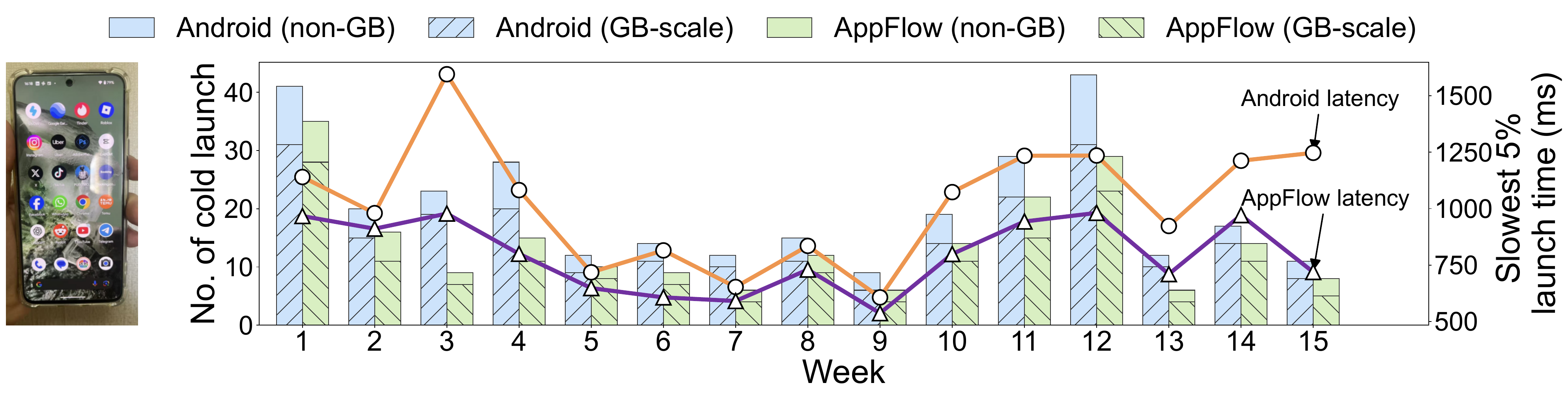} 
    \caption{\lxc{AppFlow's performance in a 100-day test.}}
\label{fig:exp:every_day}
\vspace{-2mm}
\end{figure}

\begin{figure}[t]
  \centering
  \subfloat[In-vehicle testbed]{
    \includegraphics[height=0.1\textwidth]{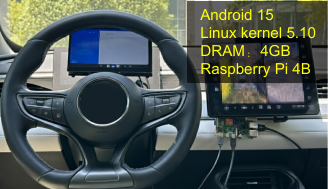}
    \label{fig:exp:car:device}
  }
  \subfloat[App-launch performance]{
    \includegraphics[height=0.12\textwidth]{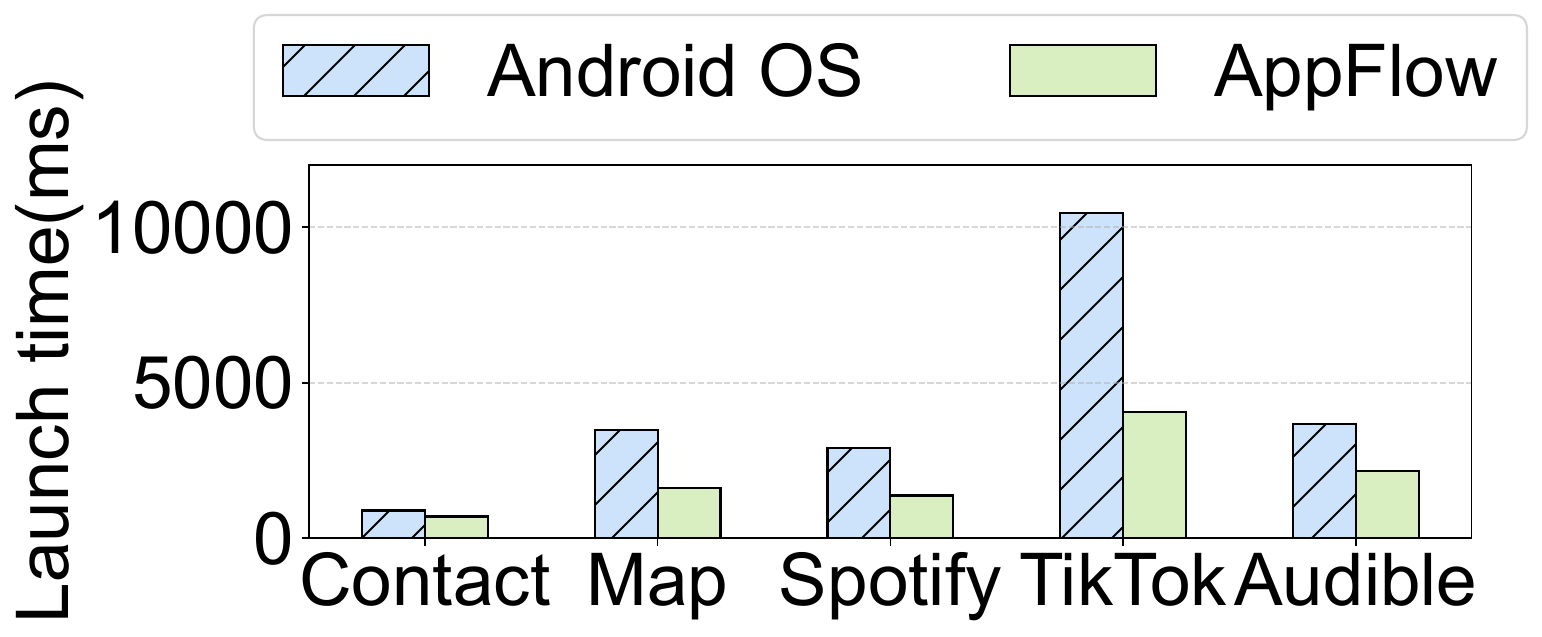}
    \label{fig:exp:car:data}
  }
  \caption{AppFlow's in-vehicle performance.}
  \label{fig:case:car}
  \vspace{-2mm}
\end{figure}

\begin{figure}[t]
  \centering
  \subfloat[On-device AGI]{
    \includegraphics[height=0.10\textwidth]{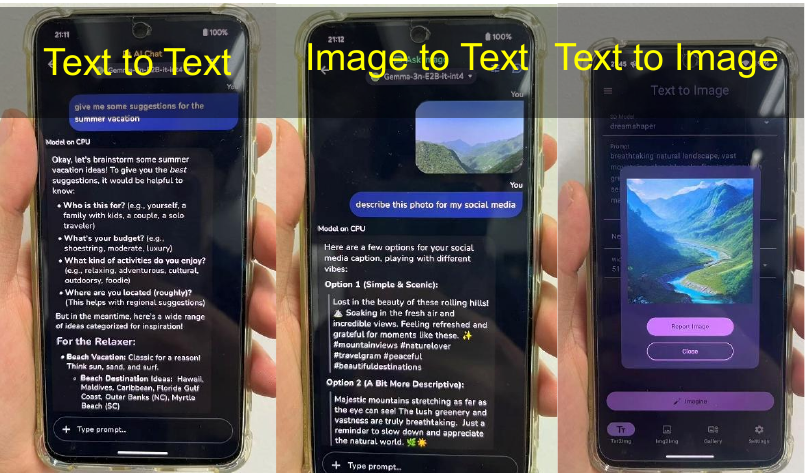}
    \label{fig:exp:agi:device}
  }
  \subfloat[Concurrency performance]{
    \includegraphics[height=0.12\textwidth]{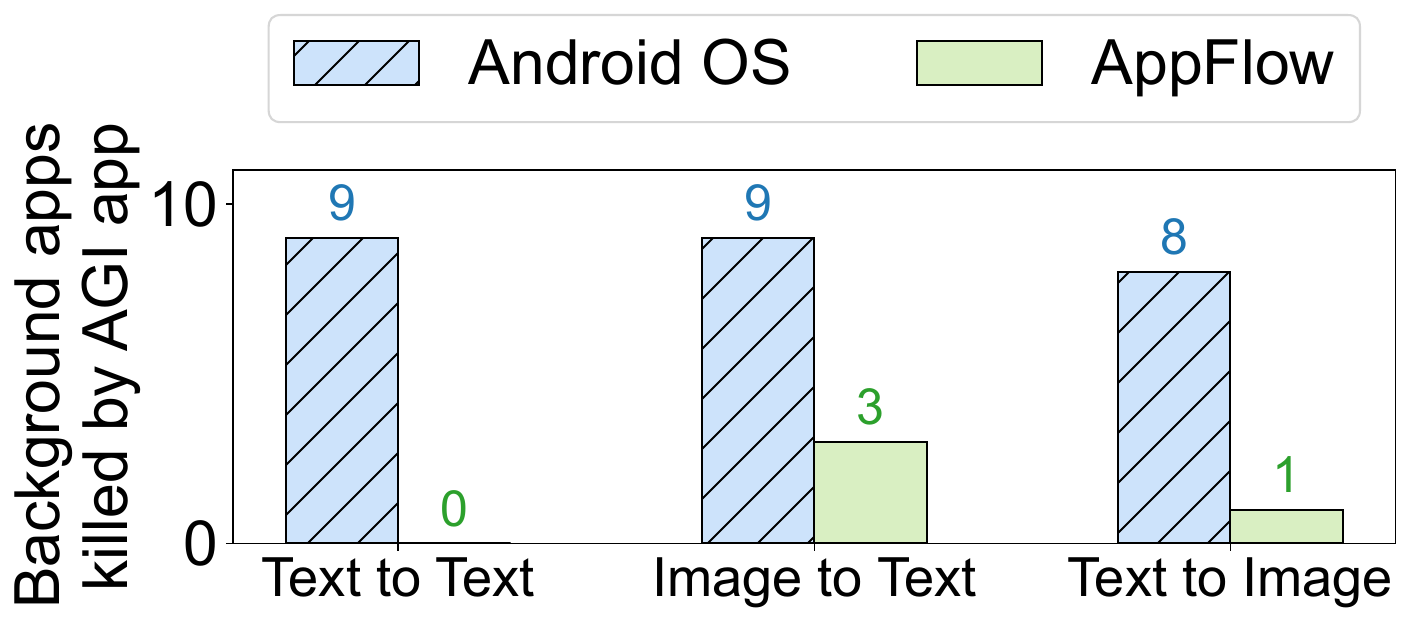}
    \label{fig:exp:agi:data}
  }
  \caption{Performance under AGI workload.}
  \label{fig:case:agi}
  \vspace{-4mm}
\end{figure}
\subsection{Case Study}
\label{sec:exp:case}
To validate \sysname’s practicality and generality, we conduct real-world case studies across diverse scenarios.

\textbf{100-day multitasking}.
In everyday use, users habitually switch between social, productivity, and media apps dozens of times a day, creating sustained memory pressure where every second of launch latency affects perceived responsiveness. 
To capture this, we test a 100-day trace of 60+ real app launches on a Google Pixel 8 (8G), comparing Android OS with \sysname. 
\lxc{As shown in \figref{fig:exp:every_day}, GB-scale apps accounted for 75.7\% of all cold launches experienced by the user. \sysname cuts average cold-launch latency by 23\% and GB-scale cold-relaunch counts by 31.6\%, demonstrating \sysname's consistent, long-term gains in everyday multitasking. For example, in the third week, while the user is engaging in a travel planning task, the Android system triggers cold launches for TikTok and Rednote, resulting in frequent stuttering. In contrast, with AppFlow, these apps are hot-launched, ensuring a smooth user experience.}

\textbf{In-vehicle systems.}
When a driver starts the car, the infotainment system boots with the ignition, clearing all processes and forcing apps such as navigation or music to launch cold, making cold launch latency directly noticeable. 
We deployed the latest Android OS 15 and \sysname on BYD Seal vehicle, equipped with a Raspberry Pi 4B with 4 GB memory (smaller than typical smartphones), powered directly by the vehicle ignition (\figref{fig:exp:car:device}), ensuring every launch was a true cold launch. 
We then measured the cold-launch latency of five popular apps, 
\eg
TikTok-1.1GB~\cite{TikTok2025}(\figref{fig:exp:car:data}). 
\sysname reduces cold-launch latency by 53.4\% (4.3s→2.0s), compared to Android OS. \lxc{This demonstrates significant benefits in resource-constrained automotive environments, where system responsiveness directly impacts the user experience upon every vehicle startup.}

\textbf{On-device AGI.} On-device AGI apps~\cite{ShiftHackZ2025Stable, GoogleAIEdgeGallery2025} represent an emerging trend, offering offline, privacy-preserving generative intelligence powered by on-device V/LLM models.
However, their massive memory footprint severely challenges multitasking. 
For example, when a user drafts an Instagram post with an image-to-text app and then switches back, Instagram is often killed, disrupting the workflow. 
To reproduce this scenario, we launched an AGI app followed by background switching on a Google Pixel 8(8G) (\figref{fig:case:agi}). 
Android OS killed 8$\sim$9 background apps after AGI use, while \sysname limited this to 0$\sim$3, boosting multitask survivability from 66.6\% to 100\%. 
\lxc{This ensures that after using the on-device AGI service, users can quickly switch to other apps without waiting through a long cold launch.}

\textbf{Summary.}
Across long-term (100-day) daily use, ignition-driven in-vehicle automotive systems, and emerging on-device AGI workloads, \sysname consistently reduces launch latency and prevents cold relaunches, demonstrating robustness across diverse mobile environments and its potential as a foundation for next-generation mobile intelligence.

\section{Related Work}
\subsection{Mobile App Launch Optimization}
App launch time is a key user-experience metric~\cite{hort2021survey,li2022smartphone,nielsen1994usability}. 
Prior work tackles it by either \textit{caching more apps for hot launch}~\cite{liang2025ariadne,parate2013practical,yan2012fast,joo2011fast} (via swap space~\cite{zhu2017smartswap,lebeck2020end}, \textit{data compression}~\cite{liang2025ariadne,li2025archer}), or \textit{speeding up cold launch}~\cite{liang2020acclaim,li2025pmr} (via file preloading~\cite{yan2012fast,garg2024crossprefetch,ryu2023fast} and faster reclamation~\cite{li2025pmr,liang2020acclaim}). 
Yet with GB-scale working sets~\cite{androidauthority_ram2025}, caching fails to cover most apps, making cold launches unavoidable~\cite{huang2024more}, while cold-launch accelerators waste resources by competing for scarce I/O and memory~\cite{son2021asap,garg2024crossprefetch}.
\sysname is the first to accelerate GB-scale cold launches while preserving background survivability, by tightly coupling preloading, reclamation, and app killing mechanisms.

\subsection{Preloading on Mobile Devices}
\label{sec:related:preload}
Preloading falls into two types, \ie \textit{before-launch}~\cite{parate2013practical,lee2016context,lee2017cas} and \textit{during-launch preloading}~\cite{joo2011fast,ryu2023fast,garg2024crossprefetch} based on the \textit{timing} of the preload.
\textit{Before-launch schemes} predict the next app a user will open and stage its working set in RAM~\cite{parate2013practical,lee2016context,lee2017cas}. For instance, MAPLE~\cite{khaokaew2024maple} prompts an LLM with user context to make that guess, yet its 52\% hit rate means every miss wastes hundreds of megabytes and leaves cold-start latency untouched, an unacceptable trade-off for GB-scale large apps.
\textit{During-launch schemes} skip prediction altogether. 
Once the user taps an icon, they stream the app’s needed pages over idle I/O bandwidth~\cite{joo2011fast,ryu2023fast,garg2024crossprefetch}. 
Identifying those pages is reliable, such as Paralfetch~\cite{ryu2023fast} covers 97.2\% of them via offline profiling.
\sysname adopts the on-demand during-launch approach, but goes further.
It selectively preloads files to boost launch speed with modest memory overhead, while cooperating seamlessly with memory-reclamation mechanisms.

\subsection{Mobile Memory Reclamation}
\label{sec:related:reclaim}
Memory reclamation is triggered when free RAM falls below a threshold~\cite{saxena2010flashvm}. 
Prior work follows two directions: deciding \textit{what/when} to reclaim (\eg Acclaim~\cite{liang2020acclaim} prioritizes background apps~\cite{zhu2017smartswap,guo2015mars}) and shortening \textit{how long} reclamation takes (\eg SWAM compresses pages in DRAM instead of writing to Flash~\cite{lim2023swam,li2025pmr,huang2024more}). 
Our work instead focuses on \textit{preloading-aware target selection}, as evicting the wrong pages can negate the benefits of preloading.

\section{Conclusion}
To address the rising cold-launch latency of GB-scale large mobile apps under constrained memory and I/O, \sysname aims to guarantee that both foreground and background launches complete within 1s. 
It achieves this through three coordinated modules: a Selective File Preloader for latency-critical files, an Adaptive Memory Reclaimer that protects preloaded data, and a Context-Aware App Killer. 
Together, these modules reduce stalls, reclaim memory efficiently, and sustain multitasking. Evaluations show \sysname cuts cold launch latency significantly while keeping relaunches within 1s even under heavy workloads.
Across 100-day daily traces, in-vehicle systems, and emerging on-device AGI, \sysname consistently reduces latency and prevents cold relaunches, showing robustness in diverse real-world scenarios.
\lxc{While \sysnameposs gains are limited for small, low-I/O apps (\eg Calendar), it introduces no regression. In the future, on-device apps (notably local AI agents) will continue to grow in size and I/O footprints, while DRAM capacity remains constrained by rising DRAM prices; in this regime, \sysname is well positioned to address the increasingly dominant I/O-latency bottleneck.}
In future work, we will incorporate user-behavior prediction to shift from app-wise to activity-wise preloading and reclamation, further accelerating launches.

\section{Acknowledgement}
This work was partially supported by the National Key R\&D Program of China (No.2024YFB4505502), the National Science Fund for Distinguished Young Scholars (62025205) and the National Natural Science Foundation of China (62522215, 62532009, 62472354)

\newpage
\bibliography{acmart}

@inproceedings{huang2024more,
  title={More Apps, Faster Hot-Launch on Mobile Devices via Fore/Background-aware GC-Swap Co-design},
  author={Huang, Jiacheng and Zhang, Yunmo and Qiu, Junqiao and Liang, Yu and Ausavarungnirun, Rachata and Li, Qingan and Xue, Chun Jason},
  booktitle={Proceedings of the 29th ACM International Conference on Architectural Support for Programming Languages and Operating Systems, Volume 3},
  pages={654--670},
  year={2024}
}

@misc{trendforce_dram_spot_2025,
  author       = {{TrendForce}},
  title        = {DRAM spot price trends},
  year         = {2025},
  month        = {July},
  day          = {25},
  howpublished = {TrendForce DRAM Price Trends (DRAM Spot Price)},
  url          = {https://www.trendforce.com/price/dram/dram_spot},
  note         = {Last update: 2025‑07‑25 Accessed: 2025‑07‑27}
}

@misc{android15_about,
  title        = {Android15 — About the Android platform version 15},
  author       = {{Google Android Developers}},
  year         = {2025},
  howpublished = {\url{https://developer.android.com/about/versions/15}},
  note         = {Accessed: 2025‑08‑01},
}

@misc{instagram,
  author       = {{Instagram}},
  title        = {Instagram},
  howpublished = {\url{https://www.instagram.com/}},
  note         = {Accessed: 2025-08-01}
}

@misc{android_logcat,
  title        = {Logcat | Android Developers},
  author       = {{Google Android Developers}},
  year         = {2025},
  howpublished = {\url{https://developer.android.com/tools/logcat}},
  note         = {Accessed: 2025-08-01},
}

@misc{android_adb,
  title        = {Android Debug Bridge (adb)},
  author       = {{Google Android Developers}},
  year         = {2025},
  howpublished = {\url{https://developer.android.com/tools/adb}},
  note         = {Accessed: 2025-08-01},
}

@article{khaokaew2024maple,
  title={Maple: Mobile app prediction leveraging large language model embeddings},
  author={Khaokaew, Yonchanok and Xue, Hao and Salim, Flora D},
  journal={Proceedings of the ACM on Interactive, Mobile, Wearable and Ubiquitous Technologies},
  volume={8},
  number={1},
  pages={1--25},
  year={2024},
  publisher={ACM New York, NY, USA}
}

@misc{carbone2022_dram_price,
  author       = {Carbone, James},
  title        = {DRAM price increases will ease},
  year         = {2022},
  month        = {May},
  day          = {12},
  howpublished = {Electronics Sourcing},
  url          = {https://electronics-sourcing.com/2022/05/12/dram-price-increases-will-ease},
  note         = {Accessed: 2025‑07‑27}
}

@misc{capcut2025,
  author       = {{CapCut}},
  title        = {{CapCut Video Editor}},
  year         = {2025},
  howpublished = {\url{https://www.capcut.com/}},
  note         = {Accessed: 2025-07-29}
}

@misc{wikipedia_android_version_history,
  author       = {{Wikipedia contributors}},
  title        = {Android version history -- Wikipedia},
  year         = {2025},
  url          = {https://en.wikipedia.org/wiki/Android_version_history},
  note         = {Accessed: 2025-07-27},
  howpublished = {\url{https://en.wikipedia.org/wiki/Android_version_history}}
}

@misc{reddit_android_memory_footprint,
  author       = {{u/DeckerSU and others}},
  title        = {What normal memory footprint for an Android app?},
  year         = {2019},
  url          = {https://www.reddit.com/r/androiddev/comments/atfccg/what_normal_memory_footprint_for_an_android_app/},
  note         = {Accessed: 2025-07-27},
}

@misc{GoogleAIEdgeGallery2025,
  author = {{Google AI Edge}},
  title = {Google AI Edge Gallery: A gallery that showcases on-device ML/GenAI use cases},
  year = {2025},
  url = {https://github.com/google-ai-edge/gallery},
  note = {Accessed: 2025-07-25}
}

@misc{AndroidLmkd2025,
  author = {{Android Open Source Project}},
  title = {Low memory killer daemon},
  year = {2025},
  url = {https://source.android.com/docs/core/perf/lmkd},
  note = {Accessed: 2025-07-25}
}

@book{gorman2004understanding,
  title={Understanding the Linux virtual memory manager},
  author={Gorman, Mel},
  volume={352},
  year={2004},
  publisher={Prentice Hall Upper Saddle River}
}

@misc{TikTok2025,
  author = {{TikTok}},
  title = {TikTok - Make Your Day},
  year = {2025},
  url = {https://www.tiktok.com/},
  note = {Accessed: 2025-07-25}
}

@inproceedings{liang2020acclaim,
  title={Acclaim: Adaptive memory reclaim to improve user experience in android systems},
  author={Liang, Yu and Li, Jinheng and Ausavarungnirun, Rachata and Pan, Riwei and Shi, Liang and Kuo, Tei-Wei and Xue, Chun Jason},
  booktitle={2020 USENIX Annual Technical Conference (USENIX ATC 20)},
  pages={897--910},
  year={2020}
}

@inproceedings{lim2023swam,
  title={Swam: Revisiting swap and oomk for improving application responsiveness on mobile devices},
  author={Lim, Geunsik and Kang, Donghyun and Ham, MyungJoo and Eom, Young Ik},
  booktitle={Proceedings of the 29th Annual International Conference on Mobile Computing and Networking},
  pages={1--15},
  year={2023}
}

@inproceedings{joo2011fast,
  title={$\{$FAST$\}$: Quick application launch on $\{$Solid-State$\}$ drives},
  author={Joo, Yongsoo and Ryu, Junhee and Park, Sangsoo and Shin, Kang G},
  booktitle={9th USENIX Conference on File and Storage Technologies (FAST 11)},
  year={2011}
}

@inproceedings{ryu2023fast,
  title={Fast Application Launch on Personal $\{$Computing/Communication$\}$ Devices},
  author={Ryu, Junhee and Lee, Dongeun and Shin, Kang G and Kang, Kyungtae},
  booktitle={21st USENIX Conference on File and Storage Technologies (FAST 23)},
  pages={425--440},
  year={2023}
}

@inproceedings{son2021asap,
  title={$\{$ASAP$\}$: Fast mobile application switch via adaptive prepaging},
  author={Son, Sam and Lee, Seung Yul and Jin, Yunho and Bae, Jonghyun and Jeong, Jinkyu and Ham, Tae Jun and Lee, Jae W and Yoon, Hongil},
  booktitle={2021 USENIX Annual Technical Conference (USENIX ATC 21)},
  pages={365--380},
  year={2021}
}

@article{lee2017cas,
  title={CAS: Context-aware background application scheduling in interactive mobile systems},
  author={Lee, Joohyun and Lee, Kyunghan and Jeong, Euijin and Jo, Jaemin and Shroff, Ness B},
  journal={IEEE Journal on Selected Areas in Communications},
  volume={35},
  number={5},
  pages={1013--1029},
  year={2017},
  publisher={IEEE}
}

@inproceedings{lee2016context,
  title={Context-aware application scheduling in mobile systems: What will users do and not do next?},
  author={Lee, Joohyun and Lee, Kyunghan and Jeong, Euijin and Jo, Jaemin and Shroff, Ness B},
  booktitle={Proceedings of the 2016 ACM International Joint Conference on Pervasive and Ubiquitous Computing},
  pages={1235--1246},
  year={2016}
}

@inproceedings{parate2013practical,
  title={Practical prediction and prefetch for faster access to applications on mobile phones},
  author={Parate, Abhinav and B{\"o}hmer, Matthias and Chu, David and Ganesan, Deepak and Marlin, Benjamin M},
  booktitle={Proceedings of the 2013 ACM international joint conference on Pervasive and ubiquitous computing},
  pages={275--284},
  year={2013}
}

@article{hort2021survey,
  title={A survey of performance optimization for mobile applications},
  author={Hort, Max and Kechagia, Maria and Sarro, Federica and Harman, Mark},
  journal={IEEE Transactions on Software Engineering},
  volume={48},
  number={8},
  pages={2879--2904},
  year={2021},
  publisher={IEEE}
}

@misc{androidauthority_ram2025,
  author       = {Hadlee Simons},
  title        = {How much RAM does your phone really need in 2025?},
  year         = {2025},
  howpublished = {\url{https://www.androidauthority.com/how-much-ram-do-i-need-phone-3086661/}},
  note         = {Accessed: 2025-07-27},
  publisher    = {Android Authority}
}

@misc{appmagic_topcharts,
  author       = {{AppMagic}},
  title        = {Live Store Rankings - AppMagic},
  year         = {2025},
  howpublished = {\url{https://appmagic.rocks/top-charts/live-store-rankings}},
  note         = {Accessed: 2025-07-27}
}

@inproceedings{garg2024crossprefetch,
  title={Crossprefetch: Accelerating i/o prefetching for modern storage},
  author={Garg, Shaleen and Zhang, Jian and Pitchumani, Rekha and Parashar, Manish and Xie, Bing and Kannan, Sudarsun},
  booktitle={Proceedings of the 29th ACM International Conference on Architectural Support for Programming Languages and Operating Systems, Volume 1},
  pages={102--116},
  year={2024}
}

@misc{Jeronimo2025GlobalMemoryShortageCrisis,
  author       = {Francisco Jeronimo},
  title        = {Global Memory Shortage Crisis: Market Analysis and the Potential Impact on the Smartphone and PC Markets in 2026},
  year         = {2025},
  month        = dec,
  howpublished = {\url{https://www.idc.com/resource-center/blog/global-memory-shortage-crisis-market-analysis-and-the-potential-impact-on-the-smartphone-and-pc-markets-in-2026/}},
  note         = {IDC Blog (Markets and Trends), published 2025-12-18, accessed 2026-01-03}
}

@book{sanders2016introduction,
  title={An introduction to Unreal engine 4},
  author={Sanders, Andrew},
  year={2016},
  publisher={AK Peters/CRC Press}
}

@misc{pubgmobile,
  author       = {{PUBG Corporation}},
  title        = {{PUBG Mobile Official Website}},
  howpublished = {\url{https://www.pubgmobile.com/en-US/home.shtml}},
  note         = {Accessed: 2025-08-05}
}

@misc{collective-ai-appstore,
  title        = {Collective AI: Secure Offline AI Assistant},
  author       = {{Whitespace Global}},
  howpublished = {Apple App Store (US)},
  year         = {2025},
  month        = oct,
  note         = {Version 1.3.0 (2025-10-02). Accessed: 2026-01-13},
  url          = {https://apps.apple.com/us/app/collective-ai/id6608983600}
}

@inproceedings{tian2020identifying,
  title={Identifying tasks from mobile app usage patterns},
  author={Tian, Yuan and Zhou, Ke and Lalmas, Mounia and Pelleg, Dan},
  booktitle={Proceedings of the 43rd international ACM SIGIR conference on research and development in information retrieval},
  pages={2357--2366},
  year={2020}
}

@article{guo2015mars,
  title={$ mars $: Mobile application relaunching speed-up through flash-aware page swapping},
  author={Guo, Weichao and Chen, Kang and Feng, Huan and Wu, Yongwei and Zhang, Rui and Zheng, Weimin},
  journal={IEEE Transactions on Computers},
  volume={65},
  number={3},
  pages={916--928},
  year={2015},
  publisher={IEEE}
}

@misc{ShiftHackZ2025Stable,
  author       = {ShiftHackZ},
  title        = {{Stable-Diffusion-Android}: Stable Diffusion AI Client (Android)},
  howpublished = {\url{https://github.com/ShiftHackZ/Stable-Diffusion-Android}},
  note         = {Accessed: 2025-09-01},
  year         = {2025}
}

@inproceedings{saxena2010flashvm,
  title={$\{$FlashVM$\}$: Virtual Memory Management on Flash},
  author={Saxena, Mohit and Swift, Michael M},
  booktitle={2010 USENIX Annual Technical Conference (USENIX ATC 10)},
  year={2010}
}

@inproceedings{yan2012fast,
  title={Fast app launching for mobile devices using predictive user context},
  author={Yan, Tingxin and Chu, David and Ganesan, Deepak and Kansal, Aman and Liu, Jie},
  booktitle={Proceedings of the 10th international conference on Mobile systems, applications, and services},
  pages={113--126},
  year={2012}
}

@inproceedings{lebeck2020end,
  title={End the senseless killing: Improving memory management for mobile operating systems},
  author={Lebeck, Niel and Krishnamurthy, Arvind and Levy, Henry M and Zhang, Irene},
  booktitle={2020 USENIX Annual Technical Conference (USENIX ATC 20)},
  pages={873--887},
  year={2020}
}

@article{kim2017application,
  title={Application-aware swapping for mobile systems},
  author={Kim, Sang-Hoon and Jeong, Jinkyu and Kim, Jin-Soo},
  journal={ACM Transactions on Embedded Computing Systems (TECS)},
  volume={16},
  number={5s},
  pages={1--19},
  year={2017},
  publisher={ACM New York, NY, USA}
}

@inproceedings{li2025pmr,
  title={$\{$PMR$\}$: Fast Application Response via Parallel Memory Reclaim on Mobile Devices},
  author={Li, Wentong and Chang, Li-Pin and Mao, Yu and Shi, Liang},
  booktitle={2025 USENIX Annual Technical Conference (USENIX ATC 25)},
  pages={1569--1584},
  year={2025}
}

@inproceedings{li2024elasticzram,
  title={Elasticzram: Revisiting zram for swapping on mobile devices},
  author={Li, Wentong and Yu, Dingcui and Song, Yunpeng and Luo, Longfei and Shi, Liang},
  booktitle={Proceedings of the 61st ACM/IEEE Design Automation Conference},
  pages={1--6},
  year={2024}
}

@inproceedings{liang2022cachesifter,
  title={$\{$CacheSifter$\}$: Sifting cache files for boosted mobile performance and lifetime},
  author={Liang, Yu and Pan, Riwei and Ren, Tianyu and Cui, Yufei and Ausavarungnirun, Rachata and Chen, Xianzhang and Li, Changlong and Kuo, Tei-Wei and Xue, Chun Jason},
  booktitle={20th USENIX Conference on File and Storage Technologies (FAST 22)},
  pages={445--459},
  year={2022}
}

@book{nielsen1994usability,
  title={Usability engineering},
  author={Nielsen, Jakob},
  year={1994},
  publisher={Morgan Kaufmann}
}

@misc{buildfire2025,
  author = {BuildFire},
  title = {Mobile App Download Statistics \& Usage Statistics (2025)},
  year = {2021},
  url = {https://buildfire.com/app-statistics/},
  note = {Accessed: 2025-06-29}
}

@inproceedings{shin2012understanding,
  title={Understanding and prediction of mobile application usage for smart phones},
  author={Shin, Choonsung and Hong, Jin-Hyuk and Dey, Anind K},
  booktitle={proceedings of the 2012 ACM conference on ubiquitous computing},
  pages={173--182},
  year={2012}
}

@article{li2022smartphone,
  title={Smartphone app usage analysis: datasets, methods, and applications},
  author={Li, Tong and Xia, Tong and Wang, Huandong and Tu, Zhen and Tarkoma, Sasu and Han, Zhu and Hui, Pan},
  journal={IEEE Communications Surveys \& Tutorials},
  volume={24},
  number={2},
  pages={937--966},
  year={2022},
  publisher={IEEE}
}

@inproceedings{zhu2017smartswap,
  title={SmartSwap: High-performance and user experience friendly swapping in mobile systems},
  author={Zhu, Xiao and Liu, Duo and Zhong, Kan and Ren, Jinting and Li, Tao},
  booktitle={Proceedings of the 54th Annual Design Automation Conference 2017},
  pages={1--6},
  year={2017}
}

@inproceedings{liang2025ariadne,
  title={Ariadne: A hotness-aware and size-adaptive compressed swap technique for fast application relaunch and reduced cpu usage on mobile devices},
  author={Liang, Yu and Shen, Aofeng and Xue, Chun Jason and Pan, Riwei and Mao, Haiyu and Ghiasi, Nika Mansouri and Jiang, Qingcai and Nadig, Rakesh and Li, Lei and Ausavarungnirun, Rachata and others},
  booktitle={2025 IEEE International Symposium on High Performance Computer Architecture (HPCA)},
  pages={1588--1602},
  year={2025},
  organization={IEEE}
}

@inproceedings{li2025archer,
  title={Archer: Adaptive Memory Compression with $\{$Page-Association-Rule$\}$ Awareness for $\{$High-Speed$\}$ Response of Mobile Devices},
  author={Li, Changlong and Zhu, Zongwei and Wang, Chao and Liu, Fangming and Xu, Fei and Sha, Edwin H-M and Zhou, Xuehai},
  booktitle={23rd USENIX Conference on File and Storage Technologies (FAST 25)},
  pages={497--511},
  year={2025}
}

@inproceedings{shen2019deepapp,
  title={DeepAPP: A deep reinforcement learning framework for mobile application usage prediction},
  author={Shen, Zhihao and Yang, Kang and Du, Wan and Zhao, Xi and Zou, Jianhua},
  booktitle={Proceedings of the 17th Conference on Embedded Networked Sensor Systems},
  pages={153--165},
  year={2019}
}
\bibliographystyle{ACM-Reference-Format}
\end{document}